\documentclass{article}

\usepackage{PRIMEarxiv}
\usepackage{setspace}
\usepackage[utf8]{inputenc} 
\usepackage[T1]{fontenc}    
\usepackage{natbib}
\usepackage{hyperref}       
\usepackage{url}            
\usepackage{booktabs}       
\usepackage{amsfonts}       
\usepackage{microtype}      
\usepackage{fancyhdr}       
\usepackage{xcolor}

\definecolor{yjrev}{RGB}{0,128,128}
\usepackage{algorithm}
\usepackage{algorithmic}

\usepackage{lmodern}
\usepackage{amsmath,amssymb,amsthm}
\usepackage{bm}
\usepackage{graphicx}
\usepackage{enumitem}
\usepackage{geometry}
\usepackage{caption}
\usepackage{subcaption}
\usepackage{multirow}
\usepackage{tikz}
\usetikzlibrary{trees,shapes,positioning}
\graphicspath{{media/}}     

\title{An interpretable semiparametric Bayesian machine learning model for high-dimensional circulating epigenetic signatures of multiple myeloma}

\author{
  Saurabh Bhandari \\
  Department of Public Health Sciences \\
  University of Chicago \\
  Chicago, IL\\
  \texttt{sbhandari52@uchicago.edu} \\
  \And
  Parveen Bhatti\\
  Population Health Sciences\\
  BC Cancer Research Institute \\
  Vancouver, BC\\
  \texttt{pbhatti@bccrc.ca} \\
   \And
  Brian C.-H. Chiu\\
  Department of Public Health Sciences \\
  University of Chicago \\
  Chicago, IL\\
  \texttt{bchiu@bsd.uchicago.edu} \\
  \And
  Yuan Ji\\
  Department of Public Health Sciences \\
  University of Chicago \\
  Chicago, IL\\
  \texttt{YJi@bsd.uchicago.edu} \\
}

\begin{document}
\maketitle

\vspace{-40pt}

\begin{abstract}
Genome-wide 5-hydroxymethylcytosine (5hmC) profiles of circulating cell-free DNA (cfDNA) can reveal epigenetic changes associated with multiple myeloma (MM), but their analysis requires adjusting for known risk factors while selecting a few informative genes from thousands.  We propose semiparametric probit Bayesian Additive Regression Tree (spBART), in which the covariates enter a linear component with interpretable coefficients and the genes enter a sparse tree ensemble whose posterior inclusion probabilities select genes conditional on the covariates. The analysis selects eight candidate loci, \textit{IL1RAP}, \textit{ST5}, \textit{HERC6}, \textit{KL}, \textit{MYO1E}, \textit{ELK3}, \textit{CAPN2} and \textit{UBR4}. Selection of \textit{IL1RAP} connects the circulating signature to the \textit{IL-1/IL-6} signaling axis implicated in MM. The final model achieves an area under the receiver operating characteristic curve of 0.96 and a Brier score of 0.09 in the held-out validation set. Our simulation studies support the modeling strategy when adjustment variables act approximately linearly, but false discoveries exceed the nominal target in application-sized settings with correlated features. Overall, the proposed model identifies a compact set of circulating epigenetic signatures that are highly associated with MM.
\end{abstract}

\vspace{-20pt}

\section{Introduction}\label{Sec::1}
\vspace{-15pt}

Multiple myeloma (MM) is a malignancy of clonal plasma cells in the bone marrow \citep{kyle2008ash}. In the United States, an estimated 36{,}110 people were diagnosed with MM and 12{,}030 died of it in 2025 \citep{siegel2025cancer}. Its incidence is approximately twice as high among African Americans as among European Americans \citep{waxman2010racial}. Understanding the molecular changes associated with MM is important to improve our understanding of its underlying biology.

Epigenetic alterations are one source of such molecular information. Among these alterations, 5-hydroxymethylcytosine (5hmC) is involved in DNA demethylation and has a distinct distribution across the genome, particularly in actively transcribed gene bodies \citep{ito2010role}. Circulating cell-free DNA (cfDNA) consists of DNA fragments released into the bloodstream from multiple cellular sources, providing a minimally invasive means of studying molecular changes associated with disease. Genome-wide profiling of cfDNA-derived 5hmC can therefore identify disease-associated epigenetic patterns without requiring tissue sampling \citep{han2016highly}. In MM, an important question is which gene-body features contribute to these patterns and how their associations should be interpreted when the circulating DNA comes from several cellular sources.

Previous work has shown differences in cfDNA-derived 5hmC between MM and its precursor conditions, including population-specific pathways \citep{chiu2022genome}. These findings motivate an examination of the joint information carried by a smaller set of features after adjustment for demographic and lifestyle variables. A locus may be selected because it contributes information beyond the other measured features, because it represents a correlated group of features, or because it captures differences in the participants or sample collection. Distinguishing these possibilities is important for deciding which findings warrant further biological investigation.

Answering these questions poses a high-dimensional statistical problem. In our application, we observe 5hmC at 14{,}147 genes in 869 participants pooled from two studies, together with age, sex, race and body mass index (BMI), which are established or suspected MM risk factors and potential confounders.  Prediction accuracy alone is not the only objective as scientists are also interested in the form of interpretable inference models. A model that predicts well but cannot answer these questions offers little guidance for such decisions \citep{rudin2019stop}. We therefore seek inference that is interpretable in two specific senses: i) direct estimates, with credible intervals, of the associations of the prespecified covariates, and ii) a sparse set of candidate genes, each with a measure of how strongly and how consistently the data support it.

Existing tools meet only part of this need. Penalized regression, such as the lasso and the elastic net \citep{tibshirani1996regression,zou2005regularization}, provides coefficients and sparse selection but assumes additive linear gene effects. Bayesian additive regression trees (BART) \citep{chipman2010bart} capture nonlinear effects and interactions without specifying them in advance and quantify posterior uncertainty in estimated quantities. When all predictors enter the tree ensemble, however, a few covariates compete with thousands of genes for splits. A covariate's effect may then be distributed across multiple splits and shrunk toward zero, while genes correlated with a strong covariate may be selected in its place. Our simulations show both problems. With sparse Dirichlet BART \citep{linero2018bayesian}, average marginal covariate effects recover about 15\% of their true size, their 95\% intervals cover the truth in 13--23\% of datasets, and null genes correlated with age are selected (Section~\ref{Sec::31}). We address these problems with a semiparametric probit model, spBART, in which the covariates enter the model as a linear component, and the 5hmC features, together with a study indicator, enter the sparse tree ensemble. The linear component keeps the covariates in the model, estimates their associations with uncertainty, and lets genes be selected conditional on them. The tree ensemble retains the flexibility of BART for the molecular signal, and a sparsity-inducing prior on its splitting variables yields posterior inclusion probabilities (PIPs). We select genes by applying the Bayesian false discovery rate (FDR) rule \citep{muller2006fdr} to the PIPs. We also examine how consistently each gene is selected across cross-validation folds, independent Markov chain Monte Carlo (MCMC) runs and an analysis of BC case-control alone.

The proposed spBART draws on two lines of work that extend BART \citep[see][for reviews]{tan2019bayesian,hill2020bayesian}. The first concerns variable selection. \citet{chipman2010bart} ranked predictors by how often they are used in splits, \citet{bleich2014variable} compared these inclusion proportions with permutation null distributions, and \citet{linero2018bayesian} replaced the uniform prior on splitting variables with a sparse Dirichlet prior, which concentrates splits on a few predictors and yields posterior inclusion probabilities (PIPs). \citet{linero2018smooth} combined this sparsity with soft decision trees that adapt to smooth functions and \citet{luo2024variable} reviewed and extended permutation-based selection. The second line combines trees with parametric components. In the semiparametric BART of \citet{zeldow2019semiparametric}, a linear term carries the effects of interest while BART adjusts for confounders, with a probit link for binary outcomes, and \citet{prado2025accounting} allowed the two components to share covariates. Bayesian causal forests \citep{hahn2020bayesian} and \texttt{stan4bart} \citep{dorie2022stan} model treatment effects alongside BART, and VCBART \citep{deshpande2024vcbart} lets the coefficients of a linear model vary through trees. These semiparametric models target low-dimensional treatment or regression effects and place no sparsity prior on the tree component, whereas the sparse BART methods provide no separate parametric component for adjustment variables. The proposed spBART combines the two and reverses the usual division of roles as the linear component holds the adjustment variables, and the sparse tree ensemble holds the high-dimensional features whose selection is the aim of the analysis.

We apply spBART to cfDNA 5hmC profiles from 869 participants (575 cases and 294 controls) in the University of Chicago MM Epidemiology (MMEP) study and the British Columbia Case-control (BC case-control) study. Pooling expands the sample and its racial diversity, but all controls come from BC, so study membership is partly aligned with case status. We include a study indicator in the tree ensemble and examine discrimination within BC, where both cases and controls are available. In the held-out validation set, the final model achieves an area under the receiver operating characteristic curve (AUC) of 0.96. Of the eight selected loci, \textit{IL1RAP}, which encodes a co-receptor in the IL-1 signaling pathway implicated in MM \citep{donovan1998contrast,lust2009induction}.

The remainder of the article is organized as follows. Section~\ref{Sec::2} presents the data, the study design, and the proposed model. Section~\ref{Sec::prior} specifies the prior distributions and describes posterior computation, gene selection, and the evaluation and sensitivity analyses. Section~\ref{Sec::4} applies spBART to the MM data, and Section~\ref{Sec::3} reports simulation studies that assess the model and the selection procedure across various settings. Section~\ref{Sec::5} concludes with a discussion.

\vspace{-10pt}  

\section{Data and model}\label{Sec::2}
\vspace{-15pt}

\subsection{Data and study design}\label{Sec::21} \vspace{-10pt}

We pool participants from two studies. The University of Chicago MM Epidemiology (MMEP) study contributes 293 MM cases, and the British Columbia Case-control (BC case-control) study contributes 282 cases and 294 controls, giving $N=869$ participants, of whom 575 are cases (Table~\ref{Table21}). Both studies record age, sex, race, and body mass index (BMI). We restrict the analysis to participants recorded as European American (EA) or African American (AA) and dichotomize BMI at 25~kg/m$^2$. In MMEP, age and BMI refer to the time of diagnosis, whereas in BC case-control, BMI is computed from the current height and weight recorded by the study.

For each participant, cfDNA 5hmC was profiled with the 5hmC-Seal method \citep{han2016highly} and summarized as read counts in 19{,}100 gene bodies. We retain the genes with at least 10 counts in at least 95\% of samples, which leaves 14{,}147 genes, and apply the variance-stabilizing transformation of \texttt{DESeq2} \citep{love2014moderated}.

All 294 controls come from BC case-control, and MMEP contributes only cases. Study membership is therefore partly aligned with case status, and differences in recruitment and study populations may contribute to the observed associations and predictive performance. Race is nearly confounded with study as well: 88 of the 93 AA participants are MMEP cases, and BC case-control includes only five. Adjusting for study membership cannot resolve the absence of MMEP controls. We include a study indicator in our model, interpret the race coefficient only as an adjustment, and assess discrimination within BC case-control, where both cases and controls are available (Section~\ref{Sec::23}). Although these steps cannot fully separate study-related differences from disease-associated signals, they allow us to examine the robustness of our findings within the available data.

Comparing cases with controls from a separate source has precedent in MM. The genome-wide association studies that identified several MM susceptibility loci compared cases recruited through clinical trials with controls from separate population studies \citep{broderick2012common, mitchell2016genome}. Germline genotypes, however, are fixed, and their measurement depends little on how samples are handled. cfDNA 5hmC can change with disease and with sample collection and processing, so this precedent carries over to our setting only in part. The Breast Cancer Association Consortium pooled breast cancer cases from 34 studies but restricted its case--control analyses to the 12 population-based studies that had their own controls \citep{yang2011associations}, and our analyses within BC case-control follow the same principle.

We split the participants into a development set ($N_{\mathrm{Dev}}=500$) and a held-out validation set ($N_{\mathrm{Val}}=369$), stratified by MM status, so that both sets retain about 66\% cases and include participants from both studies (Table~\ref{Table21}). This provides a held-out internal evaluation within the pooled study population. Gene screening, gene selection, and model fitting use only the development set, and the validation outcomes enter only the calculation of predictive performance.

Because 14{,}147 genes are many relative to 500 participants, we first screen the genes in the development set in two stages. We fit a separate probit regression of MM status on each gene, adjusting for age and sex, and keep the genes whose coefficient has a nominal $p$-value below 0.05. We then fit Gaussian mixtures with two to four components to the absolute estimated coefficients $|\widehat{b}_d|$ of the kept genes using \texttt{mclust} \citep{scrucca2016mclust}, choose the number of components by the Bayesian information criterion (BIC), and retain the genes in the component with the largest mean. 

Screening, and the fold-wise selection described in Section~\ref{Sec::242}, adjust only for covariates that are measured in the same way in both studies, cannot be affected by MM, and do not largely track study membership. Age and sex meet these conditions. Race does not, because 99.1\% of BC case-control participants are EA, whereas 30.0\% of the MMEP participants, all of them cases, are AA (Table~\ref{Table21}). BMI does not, because it was measured at diagnosis in MMEP but is current BMI in BC case-control, which for cases follows diagnosis and may reflect disease-related weight loss. Race and BMI are added as covariates only in the refit, which is fitted after the candidate genes are selected (Section~\ref{Sec::43}). Screening does not adjust for the study indicator, which enters the tree ensemble of every later fit.

We write the data as $\mathcal{D}=\{(Y_i,\mathbf{X}_i,\mathbf{Z}_i,S_i):i=1,\ldots,N\}$, where $Y_i=1$ for an MM case and $Y_i=0$ for a control, $\mathbf{X}_i=(X_{i1},\ldots,X_{iD})^\top$ contains the transformed 5hmC levels of the $D$ candidate genes, $\mathbf{Z}_i=(Z_{i1},\ldots,Z_{iJ})^\top$ contains the $J=4$ covariates, and $S_i$ indicates study membership,
\begin{equation}\label{Eq::studyIndicator}
    S_i =
    \left\{
    \begin{array}{ll}
        1 & \mbox{for the MMEP study}, \\
        0 & \mbox{for the BC case-control study}.
    \end{array}
    \right.
\end{equation}
We code age in years and sex, race and BMI as indicators of male sex, EA race and BMI $\geq 25$~kg/m$^2$. The tree ensemble below uses $\widetilde{\mathbf{X}}_i=(\mathbf{X}_i^\top,S_i)^\top$, which has $D^\ast=D+1$ components.

\begin{table}[ht]
\centering
\caption{Participant characteristics by study and by analysis set. The development and validation sets split the pooled data, stratified by MM status. Continuous variables are summarized as mean (standard deviation) and categorical variables as count (percentage). EA, European American; AA, African American.}
\label{Table21}
\renewcommand{\arraystretch}{1.2}
\resizebox{\textwidth}{!}{%
\begin{tabular}{llccccc}
\toprule
 & & \multicolumn{3}{c}{\textbf{By study}} & \multicolumn{2}{c}{\textbf{By analysis set}} \\
\cmidrule(lr){3-5} \cmidrule(lr){6-7}
\textbf{Characteristic} & & \textbf{MMEP} & \textbf{BC case-control} & \textbf{Pooled} & \textbf{Development} & \textbf{Validation} \\
\midrule
$N$ & & 293 & 576 & 869 & 500 & 369 \\
\midrule
\multirow{2}{*}{MM status} & Case & 293 (100.0\%) & 282 (49.0\%) & 575 (66.2\%) & 331 (66.2\%) & 244 (66.1\%) \\
 & Control & 0 (0.0\%) & 294 (51.0\%) & 294 (33.8\%) & 169 (33.8\%) & 125 (33.9\%) \\
\midrule
Age (years) & mean (SD) & 61.8 (9.9) & 69.1 (8.2) & 66.6 (9.5) & 66.9 (9.6) & 66.2 (9.2) \\
\midrule
\multirow{2}{*}{Sex} & Male & 147 (50.2\%) & 333 (57.8\%) & 480 (55.2\%) & 262 (52.4\%) & 218 (59.1\%) \\
 & Female & 146 (49.8\%) & 243 (42.2\%) & 389 (44.8\%) & 238 (47.6\%) & 151 (40.9\%) \\
\midrule
\multirow{2}{*}{BMI} & $<25$ kg/m$^2$ & 84 (28.7\%) & 192 (33.3\%) & 276 (31.8\%) & 159 (31.8\%) & 117 (31.7\%) \\
 & $\geq 25$ kg/m$^2$ & 209 (71.3\%) & 384 (66.7\%) & 593 (68.2\%) & 341 (68.2\%) & 252 (68.3\%) \\
\midrule
\multirow{2}{*}{Race} & EA & 205 (70.0\%) & 571 (99.1\%) & 776 (89.3\%) & 446 (89.2\%) & 330 (89.4\%) \\
 & AA & 88 (30.0\%) & 5 (0.9\%) & 93 (10.7\%) & 54 (10.8\%) & 39 (10.6\%) \\
\midrule
\multirow{2}{*}{Study} & MMEP & -- & -- & 293 (33.7\%) & 163 (32.6\%) & 130 (35.2\%) \\
 & BC case-control & -- & -- & 576 (66.3\%) & 337 (67.4\%) & 239 (64.8\%) \\
\bottomrule
\end{tabular}}
\end{table}

\vspace{-10pt}

\subsection{Semiparametric probit BART model}\label{Sec::22}\vspace{-10pt}

We model MM status with a probit model whose latent mean has two parts, a linear function of the covariates and a sum of regression trees in the genes and the study indicator,
\begin{equation}\label{eq:spbart_model}
    Y_i = \mathbb{I}(U_i > 0), \qquad
    U_i = f(\widetilde{\mathbf{X}}_i) + \mathbf{Z}_i^\top \boldsymbol{\beta} + \varepsilon_i,
    \qquad \varepsilon_i \sim \mathcal{N}(0,1).
\end{equation}
Model~\eqref{eq:spbart_model} implies $\Pr(Y_i=1\mid\widetilde{\mathbf{X}}_i,\mathbf{Z}_i)=\Phi\{f(\widetilde{\mathbf{X}}_i)+\mathbf{Z}_i^\top\boldsymbol{\beta}\}$.

The function $f$ is a sum of $T$ regression trees \citep{chipman2010bart}, $f(\widetilde{\mathbf{x}})=\sum_{t=1}^{T}g(\widetilde{\mathbf{x}};\mathcal{T}_t,\boldsymbol{\mu}_t)$. Each tree $\mathcal{T}_t$ partitions the predictor space by binary splits of the form $\widetilde{x}_j\le s$, and $g(\widetilde{\mathbf{x}};\mathcal{T}_t,\boldsymbol{\mu}_t)$ returns the value $\mu_{t\ell}$ of the terminal node $\ell$ that contains $\widetilde{\mathbf{x}}$. Each tree adds a small step to the latent mean, and the sum of many shallow trees represents nonlinear effects and interactions. 

The linear part of~\eqref{eq:spbart_model} contains age, sex, race, and BMI. These are established or suspected risk factors for MM and potential confounders, which we specify in advance, and we want to report the association of each with MM status together with its uncertainty. The genes, together with the study indicator discussed below, enter the tree ensemble. We do not know in advance which genes are associated with MM or in what form, and the sum of trees learns both from the data. Because age, sex, race, and BMI enter only the linear part, the splits are left to the genes and the study indicator, and the genes are selected conditional on the linear covariate associations (Section~\ref{Sec::242} gives the covariates used at each step of the application).

We place the study indicator in the tree ensemble for two reasons. First, because MMEP contributes only cases, $S_i=1$ implies $Y_i=1$ in the pooled data. Therefore, a linear coefficient for $S_i$ would face quasi-complete separation, as the likelihood is maximized only as the coefficient tends to infinity, so its estimate would be determined by its prior. Second, in BART, the shrinkage prior on the leaf values limits the contribution of $S_i$, and splits on $S_i$ allow the association of a gene to differ between studies. Later, we perform a sensitivity analysis within the BC case-control data set described in Section~\ref{Sec::23} to assess the impact of $S_i$.

We use the probit link because the latent normal variable $U_i$ \citep{albert1993bayesian} reduces the updates of $f$ and $\boldsymbol{\beta}$ to Gaussian regressions, so the Bayesian backfitting algorithm of BART applies unchanged. The probit link has a drawback in a case--control study. Under a logistic model, case--control sampling changes only the intercept, and odds ratios keep their population interpretation \citep{prentice1979logistic}. The probit model has no such property. We therefore interpret $\boldsymbol{\beta}$ as associations in the sampled population, conditional on $f(\widetilde{\mathbf{X}}_i)$ and the other covariates.

\section{Prior specification and posterior inference}\label{Sec::prior}
\vspace{-15pt}

We now complete the Bayesian specification of model~\eqref{eq:spbart_model}, describe how we sample from its posterior distribution, and define the posterior summaries we use for the covariates, the genes and prediction.

\subsection{Prior specification}\label{Sec::223}\vspace{-10pt}

We follow \citet{chipman2010bart} for the tree structures. A node at depth $\delta$ splits with probability $\kappa(1+\delta)^{-\nu}$, with $\kappa=0.95$ and $\nu=2$, which favors shallow trees. The leaf values are independent, $\mu_{t\ell}\sim\mathcal{N}(0,\sigma_\mu^2)$ for $\ell=1,\ldots,L_t$, where $L_t$ is the number of terminal nodes of tree $t$. We set $\sigma_\mu=0.5/\sqrt{T}$. Because $\widetilde{\mathbf{x}}$ falls in exactly one leaf of each tree, $f(\widetilde{\mathbf{x}})$ is a priori the sum of $T$ independent $\mathcal{N}(0,\sigma_\mu^2)$ leaf values and has standard deviation $\sqrt{T}\,\sigma_\mu=0.5$ for any number of trees. About 95\% of its prior mass lies between $-1$ and $1$ on the probit scale, which corresponds to probabilities between 0.16 and 0.84 when $\mathbf{Z}_i^\top\boldsymbol{\beta}=0$. The prior therefore allows the genes and the study indicator to shift the probability of being a case moderately, and strong signals in the data can still move it close to 0 or 1. In the simulation studies, we use the \texttt{pbart} default \citep{sparapani2021nonparametric} instead, for a like-for-like comparison with Dirichlet BART (Section~\ref{Sec::3}).

To encourage the ensemble to use few predictors, we place the sparse Dirichlet prior of \citet{linero2018bayesian} on the splitting probabilities,
$(\pi_1,\ldots,\pi_{D^\ast})\sim\mathrm{Dirichlet}(\xi/D^\ast,\ldots,\xi/D^\ast)$,
where $\pi_j$ is the probability that a split uses the $j$th component of $\widetilde{\mathbf{X}}_i$. Smaller values of $\xi$ concentrate the splits on fewer predictors. We learn the degree of sparsity through $\theta=\xi/(\xi+\rho)\sim\mathrm{Beta}(0.5,1)$, which favors small $\theta$ while allowing denser splitting, and set $\rho=D^\ast$. As in the \texttt{BART} package \citep{sparapani2021nonparametric}, the splitting probabilities stay uniform during the first half of the burn-in and follow the Dirichlet prior thereafter.

For the covariate coefficients we use the conditionally conjugate prior $\boldsymbol{\beta}\sim\mathcal{N}(\mathbf{0},\sigma_\beta^2 I_J)$, where $I_J$ is the $J\times J$ identity matrix, with $\sigma_\beta^2=10$. Its 95\% prior interval for each coefficient, approximately $(-6.2,6.2)$, covers probit-scale associations far larger than those typically observed for demographic risk factors in epidemiologic studies, so the prior adds little information with four covariates and 500 participants while keeping the update of $\boldsymbol{\beta}$ in closed form. Study~6 (Appendix~\ref{App::sim6}) shows that gene selection, prediction and the coefficient estimates vary little for $\sigma_\beta^2\geq 1$.

\subsection{Posterior computation}\label{Sec::226}\vspace{-10pt}

Let $\mathcal{F}=\{(\mathcal{T}_t,\boldsymbol{\mu}_t)\}_{t=1}^{T}$ collect the trees and their leaf values. The joint posterior of $\mathcal{F}$, $\boldsymbol{\beta}$ and the latent variables $\mathbf{U}=(U_1,\ldots,U_N)^\top$ is
\begin{equation}\label{eq:joint_posterior}
\pi\big(\mathcal{F},\boldsymbol{\beta},\mathbf{U}\mid\mathbf{Y},\widetilde{\mathbf{X}},\mathbf{Z}\big)
\propto \prod_{i=1}^N \Big\{\mathbb{I}(U_i>0)^{Y_i}\,\mathbb{I}(U_i\le 0)^{1-Y_i}\,
\phi\big(U_i-f(\widetilde{\mathbf{X}}_i)-\mathbf{Z}_i^\top\boldsymbol{\beta}\big)\Big\}\,
\pi(\mathcal{F})\,\pi(\boldsymbol{\beta}),
\end{equation}
where $\phi$ is the standard normal density and $\pi(\mathcal{F})$ includes the Dirichlet prior on the splitting probabilities. Our Markov chain Monte Carlo (MCMC) sampler cycles through four updates: the latent variables from truncated normal distributions, $\boldsymbol{\beta}$ from its normal full conditional given the partial residuals $U_i-f(\widetilde{\mathbf{X}}_i)$, each tree by the Bayesian backfitting algorithm of \citet{chipman2010bart} with Metropolis--Hastings proposals, and the splitting probabilities from their Dirichlet full conditional (Appendix~\ref{App::A}). In the application, each fit uses $T=200$ trees and runs 2{,}000 burn-in iterations followed by 5{,}000 iterations thinned by 5, which gives $Q=1{,}000$ retained draws $\{(f^{(q)},\boldsymbol{\beta}^{(q)}):q=1,\ldots,Q\}$.

\subsection{Covariate associations and gene selection}\label{Sec::24}\vspace{-10pt}

The fitted model gives the two kinds of interpretable output described in Section~\ref{Sec::1}, posterior summaries of the covariate associations and posterior support for each gene.

\paragraph{Covariate associations.}\label{Sec::cov}
The linear part contains four covariates: age in years, and indicators of male sex, EA race and BMI $\geq 25$~kg/m$^2$. For each covariate $j$, we report the posterior mean of $\beta_j$, its 95\% credible interval and the posterior probability that it is positive. Each coefficient is a probit-scale association conditional on $f(\widetilde{\mathbf{X}}_i)$ and the other three covariates. Because probit coefficients are conditional and noncollapsible, they cannot be compared directly with those of a model that omits $f$ or represents the covariates through trees. In the simulation studies, we therefore also compare methods by the average marginal effect (AME) of covariate $j$ on the probability scale,
\begin{equation}\label{Eq::AME}
\mathrm{AME}_j
= \frac{1}{|\mathcal{A}|}\sum_{i\in\mathcal{A}}
\Big[\Phi\big\{\eta(\widetilde{\mathbf{X}}_i,\mathbf{z}_{i,j}^{(1)})\big\}
-\Phi\big\{\eta(\widetilde{\mathbf{X}}_i,\mathbf{z}_{i,j}^{(0)})\big\}\Big],
\end{equation}
where $\eta$ is the latent predictor of the fitted model, $\mathcal{A}$ is a set of participants, and the covariate vectors $\mathbf{z}_{i,j}^{(1)}$ and $\mathbf{z}_{i,j}^{(0)}$ differ only in covariate $j$. For a binary covariate, $\mathbf{z}_{i,j}^{(1)}$ and $\mathbf{z}_{i,j}^{(0)}$ set covariate $j$ to 1 and 0. For a continuous covariate, $\mathbf{z}_{i,j}^{(0)}$ keeps the observed value and $\mathbf{z}_{i,j}^{(1)}$ adds one standard deviation to it. Evaluating~\eqref{Eq::AME} at each posterior draw gives its posterior distribution. In the application, we take $\mathcal{A}$ to be the 369 validation participants and, for age, add 10 years instead of one standard deviation.

\paragraph{Posterior inclusion probabilities.}\label{Sec::241}
For gene $d$, let $\gamma_d^{(q)}=\mathbb{I}\{\text{gene } d \text{ is used in at least one split at draw } q\}$, and estimate its posterior inclusion probability by $\widehat{\mathrm{PIP}}_d=Q^{-1}\sum_{q=1}^{Q}\gamma_d^{(q)}$. When gene $d$ is used in no split, $f$ does not depend on $X_d$. Hence $\widehat{\mathrm{PIP}}_d$ estimates the posterior probability that $f$ depends on gene $d$, and $1-\widehat{\mathrm{PIP}}_d$ estimates the posterior probability that the gene is excluded. The sparse Dirichlet prior makes inclusion informative, because under uniform splitting a large ensemble can use many genes by chance. Exclusion and the absence of an association are different events. A gene associated with MM may be excluded because a correlated gene carries the same information, and the probabilities depend on the candidate genes and on the model, including its linear covariate terms. We therefore interpret a PIP as evidence that a gene contributes given the other predictors and the model.

\paragraph{Bayesian FDR rule and gene selection.}\label{Sec::242}
We select genes with the Bayesian false discovery rate (FDR) rule of \citet{muller2006fdr}, in which a false discovery is a selected gene that is excluded from the ensemble. If we select the genes whose PIPs are at least $\tau$, the posterior expected number of false discoveries is the sum of $1-\widehat{\mathrm{PIP}}_d$ over the selected genes, and the posterior expected FDR is
\begin{equation}\label{Eq::expected_FDR}
    \mathrm{FDR}(\tau) = \frac{\sum_{d:\widehat{\mathrm{PIP}}_d\geq\tau} \big(1-\widehat{\mathrm{PIP}}_d\big)}{\max\Big(1, \sum_{d:\widehat{\mathrm{PIP}}_d\geq\tau} 1\Big)}.
\end{equation}
To apply the rule at level $\alpha$, we rank the genes by decreasing PIP, $\widehat{\mathrm{PIP}}_{(1)}\geq\cdots\geq\widehat{\mathrm{PIP}}_{(D)}$, compute $\widehat{\mathrm{FDR}}(R)=R^{-1}\sum_{r=1}^{R}(1-\widehat{\mathrm{PIP}}_{(r)})$ for the top $R$ genes, and select the $R^\ast=\max\{R:\widehat{\mathrm{FDR}}(R)\le\alpha\}$ genes with the largest PIPs, which corresponds to the threshold $\tau=\widehat{\mathrm{PIP}}_{(R^\ast)}$. We use $\alpha=0.05$ throughout.

In the application, we apply the rule within each of the five cross-validation training sets described in Section~\ref{Sec::23}, using the PIPs $\widehat{\mathrm{PIP}}_d^{(-k)}$ from the fit that leaves out fold $k$, and take the union of the five selected sets as the candidate genes. These fold-specific fits include only age and sex in the linear part, so the candidate genes are selected conditional on age and sex (as noted in Section~\ref{Sec::21}). We then refit spBART to the whole development set with the candidate genes, the study indicator, and all four covariates, apply the rule again to its PIPs $\widehat{\mathrm{PIP}}_d^{\mathrm{full}}$, and report this refit. The posterior bound holds within each fold-specific fit but not for the union, because a null gene selected in any fold enters it. The refit PIPs are conditional on genes chosen from the same data. With few candidates, a gene may appear in every draw of the refit, so a PIP near 1 shows only that the restricted model uses it. We therefore treat the union as a set of candidate genes and measure the support for each gene by how consistently it is selected across folds, across independent chains and in the analysis of BC case-control alone.

\subsection{Prediction, evaluation and sensitivity analyses}\label{Sec::23}\vspace{-10pt}

A fit gives the predicted probability $\widehat{p}_i=Q^{-1}\sum_{q=1}^{Q}p_i^{(q)}$ for a participant with predictors $(\widetilde{\mathbf{X}}_i,\mathbf{Z}_i)$, where $p_i^{(q)}=\Phi\{f^{(q)}(\widetilde{\mathbf{X}}_i)+\mathbf{Z}_i^\top\boldsymbol{\beta}^{(q)}\}$. We partition the development set into $K=5$ folds, stratified by MM status. For fold $k$, we fit spBART to the other four folds and predict the participants in fold $k$, and we pool these out-of-fold predictions to obtain the cross-validated performance. The refit to the whole development set described in Section~\ref{Sec::242} then predicts the validation participants.

We measure discrimination by the area under the receiver operating characteristic curve (AUC) \citep{robin2011pROC} and the accuracy of the predicted probabilities by the Brier score, $\mathrm{BS}=|\mathcal{E}|^{-1}\sum_{i\in\mathcal{E}}(Y_i-\widehat{p}_i)^2$ for an evaluation set $\mathcal{E}$, which reflects both calibration and discrimination. We also compute both metrics at each posterior draw, in each held-out fold and in the validation set, and summarize them by their posterior means and 95\% credible intervals from the 2.5th and 97.5th percentiles. These intervals describe posterior uncertainty conditional on the observed evaluation participants and partitions. For each validation set, we also report the AUC of the posterior-mean predicted probabilities $\widehat{p}_i$ with a 95\% confidence interval by the method of \citet{delong1988comparing}, which we checked against 2{,}000 stratified bootstrap resamples \citep{robin2011pROC}. Unlike the credible intervals, this interval reflects the sampling variation of the evaluation participants.

Because preprocessing ignores outcomes and the validation outcomes enter only these metrics, the validation set assesses the complete pipeline of screening, selection and fitting. The cross-validated metrics are optimistic for that pipeline, because screening used all development participants before the folds were formed. We report them as a secondary summary and base our assessment of prediction on the validation set.

Because study membership is aligned with case status, we also evaluate discrimination in two analyses in which study membership is constant. First, we evaluate the pooled model on the BC case-control participants of the validation set, where study membership is constant. Second, we repeat the complete analysis with BC case-control participants only and without the study indicator, including screening, five-fold selection, the union and the refit, using a separate random split into 500 development and 76 validation participants. To check the stability of gene selection across Markov chains, we also repeat fold-wise selection, the union and the refit with four chains per fit, keeping the screened genes and the folds unchanged. Each chain has its own seed and random initialization of the latent variables and runs 10{,}000 burn-in iterations followed by 10{,}000 iterations thinned by 10, which yields 1{,}000 draws. Before pooling the chains, we assess the rank-normalized split-$\widehat{R}$ and bulk effective sample size \citep{vehtari2021rank} of the log-likelihood, the number of splits, the coefficients and the latent means of three fixed participants (Appendix~\ref{App::multichain}).

\section{Application to multiple myeloma}\label{Sec::4}\vspace{-15pt}
We apply spBART to the pooled data of Section~\ref{Sec::21}. Screening, gene selection and model fitting use only the 500 development participants, and we use the 369 validation participants only to evaluate the final model. We assess how well the model discriminates cases from controls, which genes and covariates are associated with the disease status.

\vspace{-10pt}

\subsection{Candidate genes and predictive performance}\label{Sec::41}\vspace{-10pt}
We apply the two-stage screening of Section~\ref{Sec::21} to the 14{,}147 genes in the 500 development participants. The first stage retains 8{,}974 genes whose coefficients have nominal $p$-values below 0.05. In the second stage, BIC selects four mixture components, and the component with the largest mean absolute coefficient contains 115 genes, which enter all subsequent fits.

We fit spBART as described in Sections~\ref{Sec::22}--\ref{Sec::226}. Applying the Bayesian FDR rule at $\alpha=0.05$ within each of the five development folds, whose fits adjust for age and sex, and taking the union of the selected sets gives eight candidate genes:
\begin{equation}\label{Eq::selected_genes}
\text{\textit{IL1RAP} (5), \textit{ST5} (4), \textit{HERC6} (2), \textit{MYO1E} (2), \textit{KL} (1), \textit{ELK3} (1), \textit{CAPN2} (1) and \textit{UBR4} (1),}
\end{equation}
where the number in parentheses is the number of the five folds that select the gene (Appendix~\ref{App::multichain}). We then refit spBART to the whole development set with these eight genes, the study indicator and all four covariates.

\begin{table}[ht]
\centering
\caption{Cross-validated performance of spBART in the development set ($N_{\text{Dev}} = 500$; 331 cases, 169 controls). For each fold, the fit to the other four folds predicts the participants in the fold, and we evaluate each metric at every retained draw.}
\label{Table82}
\renewcommand{\arraystretch}{1.2}
\begin{tabular}{lcccc}
\toprule
& \multicolumn{2}{c}{\textbf{AUC}} 
& \multicolumn{2}{c}{\textbf{Brier Score}} \\
\cmidrule(lr){2-3} \cmidrule(lr){4-5}
\textbf{CV Fold}  & \textbf{Posterior Mean} 
& \textbf{95\% Credible Interval} 
& \textbf{Posterior Mean} 
& \textbf{95\% Credible Interval} \\
\midrule
1 & 0.89 & (0.85, 0.93) & 0.12 & (0.11, 0.14) \\
2 & 0.93 & (0.89, 0.96) & 0.12 & (0.10, 0.13) \\
3 & 0.94 & (0.91, 0.97) & 0.11 & (0.10, 0.13) \\
4 & 0.94 & (0.90, 0.97) & 0.12 & (0.10, 0.14) \\
5 & 0.90 & (0.86, 0.92) & 0.14 & (0.12, 0.15) \\
\bottomrule
\end{tabular}
\end{table}

\paragraph{Predictive performance.}
 Across the five development folds, the AUC is 0.89--0.94 and the Brier score 0.11--0.14 (Table~\ref{Table82}). We use cross-validation mainly to select the candidate genes, and report these values for completeness. 
In the validation set, the final model has an AUC of 0.96 (95\% credible interval 0.95--0.97) and a Brier score of 0.09 (0.08--0.10) (Table~\ref{Table83}). The AUC of the posterior-mean predicted probabilities is 0.97, with DeLong 95\% confidence interval 0.95--0.98. Because the Brier score combines calibration and discrimination, assessing calibration would require comparing predicted and observed case proportions. The predicted probabilities also reflect the case--control composition of the sample, in which about two thirds of the participants are cases. Study membership alone gives a validation AUC of 0.77, because all 130 MMEP participants in the validation set are cases. Therefore, the study membership contributed to the model's discriminative performance, but only partly.

\begin{table}[ht]
\centering
\caption{Performance in the validation set ($N_{\text{Val}} = 369$; 244 cases, 125 controls) of the spBART refit to the development set. The last row gives the AUC of the posterior-mean predicted probabilities with its DeLong confidence interval, which reflects the sampling variation of the validation participants.}
\label{Table83}
\renewcommand{\arraystretch}{1.2}
\begin{tabular}{lcc}
\toprule
\textbf{Metric} & \textbf{Posterior Mean} & \textbf{95\% Credible Interval} \\
\midrule
AUC & 0.96 & (0.95, 0.97) \\
Brier Score & 0.09 & (0.08, 0.10) \\
\midrule
\textbf{Metric} & \textbf{Estimate} & \textbf{DeLong 95\% Confidence Interval} \\
AUC of posterior-mean predictions & 0.97 & (0.95, 0.98) \\
\bottomrule
\end{tabular}
\end{table}

\paragraph{Discrimination within BC case-control.}
To measure discrimination that study membership cannot explain, we use the two analyses of Section~\ref{Sec::23} in which study membership is constant (Table~\ref{Table::BConly}). First, we evaluate the pooled model, without refitting, in the 239 BC case-control participants of the validation set (114 cases and 125 controls). This is a subset of the validation set excluding all MMEP participants. Its AUC is 0.92 (95\% credible interval 0.90--0.94). The model may rank these cases above the controls through study membership, through the genes, or through both, so the difference between 0.96 and 0.92 does not isolate the contribution of study membership. The AUC of 0.92 is our most direct evidence that the genes and covariates discriminate MM cases from controls.

Second, we repeat the complete analysis with the 576 BC case-control participants alone and without the study indicator, using a separate random split into 500 development and 76 validation participants. This analysis gives a cross-validated AUC of 0.90 and a validation AUC of 0.85 (0.82--0.88), with a Brier score of 0.15 (0.14--0.17). The AUC of the posterior-mean predicted probabilities is 0.86, and with 76 validation participants its DeLong 95\% confidence interval is 0.77--0.95.

\begin{table}[ht]
\centering
\caption{Predictive performance in the pooled and BC-only analyses. Study membership alone ranks MMEP participants above BC participants. In (a), we evaluate the pooled model in its BC validation subset without refitting. In (b), we repeat screening, selection, and fitting using a separate split of the BC data into 500 development and 76 validation participants. Credible intervals describe posterior uncertainty conditional on the observed evaluation samples. The last column gives the AUC of the posterior-mean predicted probabilities with its DeLong 95\% confidence interval.}
\label{Table::BConly}
\renewcommand{\arraystretch}{1.2}
\resizebox{\textwidth}{!}{%
\begin{tabular}{llccc}
\toprule
\textbf{Model} & \textbf{Evaluation set} & \textbf{AUC} & \textbf{95\% Credible Interval} & \textbf{AUC of $\widehat{p}_i$ (95\% CI)} \\
\midrule
\multicolumn{5}{l}{\textit{Pooled analysis (500 development / 369 validation pooled participants)}} \\
Study membership only & Pooled validation set (244 cases, 125 controls) & 0.77 & -- & -- \\
Pooled spBART model & Pooled validation set (244 cases, 125 controls) & 0.96 & (0.95, 0.97) & 0.97 (0.95, 0.98) \\
\addlinespace
\multicolumn{5}{l}{\textit{(a) Within-BC evaluation of the pooled analysis (same split and model, no refitting)}} \\
Study membership only & BC case-control participants of the pooled validation set (114 cases, 125 controls) & 0.50 & -- & -- \\
Pooled spBART model & BC case-control participants of the pooled validation set (114 cases, 125 controls) & 0.92 & (0.90, 0.94) & 0.93 (0.90, 0.96) \\
\addlinespace
\multicolumn{5}{l}{\textit{(b) Separate BC-only reanalysis (MMEP removed; 500 development / 76 validation BC case-control participants)}} \\
BC-only spBART model & 5-fold cross-validation (500 BC-only development participants) & 0.90 & (0.85, 0.95) & -- \\
BC-only spBART model & BC-only validation set (37 cases, 39 controls) & 0.85 & (0.82, 0.88) & 0.86 (0.77, 0.95) \\
\bottomrule
\end{tabular}}
\end{table}

\vspace{-10pt}
\subsection{Stability of gene selection}\label{Sec::44}\vspace{-10pt}

In the refit, each of the eight genes appears in some tree at every posterior draw, so each has a PIP of 1.00, which follows from the small number of candidates. We therefore measure the support for each gene by how consistently it is selected when we change the Markov chain or the data (Table~\ref{tab:stability}). We first repeat fold-wise selection, the union and the refit with four chains per fit (Section~\ref{Sec::23}). The chains mixed well, although the log-likelihood mixed more slowly in some fold fits (Appendix~\ref{App::multichain}).

\textit{IL1RAP} is selected in every fold of every chain, and \textit{ST5} enters every chain's union and every fold when the draws of the four chains are pooled. \textit{KL} enters three of the four chain unions. The other five candidate genes, \textit{HERC6}, \textit{MYO1E}, \textit{ELK3}, \textit{CAPN2} and \textit{UBR4}, enter at most two chain unions and no pooled fold, which suggests that their selection in the single-chain analysis reflects Monte Carlo variation near the selection threshold. \textit{NF2}, which no fold of the single-chain analysis selects, enters every chain's union. Prediction changes little, with a validation AUC of 0.95 for the refit with pooled chains.

The BC-only analysis described in Section~\ref{Sec::23} changes the data instead. It removes the MMEP cases and the study indicator and repeats screening, fold-wise selection, the union and the refit in BC case-control alone (Appendix~\ref{App::BConly}). \textit{IL1RAP} and \textit{ST5} are selected again, \textit{IL1RAP} in all five folds and \textit{ST5} in two, and they are the only candidate genes that this analysis selects. The absence of the other six is consistent with how they entered the candidate set. The union keeps every gene that any fold selects, so it also admits genes with weak support, including null genes in the simulations of Section~\ref{Sec::36}. Each of the six entered the pooled union through one or two folds, and the four-chain analysis shows that their selection changes with the Markov chain even when the data are fixed. Two of them, \textit{KL} and \textit{HERC6}, also fail the BC-only screening. The BC-only union contains 19 genes. Outside the candidate set, \textit{NF2} is selected in four BC-only folds and enters every pooled chain union, whereas \textit{FKBP9} and \textit{PRKDC}, selected in five and four BC-only folds, enter no pooled union. Because \textit{IL1RAP} and \textit{ST5} are the only candidate genes selected in both the pooled and the BC-only analyses, we focus our interpretation on these two genes. As in the pooled analysis, the BC-only fold-specific fits adjust for age and sex, and the BC-only refit for all four covariates.

cfDNA comes from several cell types, so the associations of these two genes with MM status may reflect myeloma cells, the host response or both. \textit{IL1RAP} encodes a co-receptor of IL-1 signaling, which connects the circulating signature to the IL-1/IL-6 paracrine pathway implicated in myeloma proliferation and investigated through IL-1 blockade \citep{donovan1998contrast,lust2009induction}. \textit{ST5}, also known as \textit{DENND2B}, encodes a protein that binds the SH3 domain of the ABL kinase and can activate the MAP kinase ERK2 \citep{majidi1998activation}. We have found no previous report that links \textit{ST5} to MM, so its consistent selection makes it a new candidate for follow-up. One ST5 isoform has been detected in epithelial and fibroblast cell lines but not in lymphoid cell lines \citep{hubbs1999expression}, so part of the \textit{ST5} signal may come from cells other than the tumor. Protein expression in cell lines, however, need not reflect 5hmC levels in cfDNA. Within BC case-control, where study membership plays no role, the median 5hmC level of cases exceeds that of controls for both genes (Figure~\ref{fig:IL1RAP_ST5_BC}). 

\begin{table}[ht]
\centering
\caption{Stability of selection for the eight candidate genes and for three other genes discussed in the text. The columns give the number of the five folds that select each gene in the primary pooled analysis, the number of the four chains of the pooled four-chain analysis whose union contains it, and the number of the five folds that select it in the BC-only analysis. ``Not screened'' marks a gene that fails the BC-only screening. Appendix~\ref{App::multichain} and Appendix~\ref{App::BConly} give the complete results.}
\label{tab:stability}
\renewcommand{\arraystretch}{1.2}
\begin{tabular}{lccc}
\toprule
& \multicolumn{2}{c}{\textbf{Pooled analysis}} & \textbf{BC-only analysis} \\
\cmidrule(lr){2-3} \cmidrule(lr){4-4}
\textbf{Gene} & \textbf{Primary folds (of 5)} & \textbf{Chain unions (of 4)} & \textbf{Folds (of 5)} \\
\midrule
\textit{IL1RAP} & 5 & 4 & 5 \\
\textit{ST5} & 4 & 4 & 2 \\
\textit{HERC6} & 2 & 1 & not screened \\
\textit{MYO1E} & 2 & 1 & 0 \\
\textit{KL} & 1 & 3 & not screened \\
\textit{ELK3} & 1 & 1 & 0 \\
\textit{CAPN2} & 1 & 1 & 0 \\
\textit{UBR4} & 1 & 2 & 0 \\
\addlinespace
\textit{NF2} & 0 & 4 & 4 \\
\textit{FKBP9} & 0 & 0 & 5 \\
\textit{PRKDC} & 0 & 0 & 4 \\
\bottomrule
\end{tabular}
\end{table}

\vspace{-10pt}
\subsection{Covariate associations}\label{Sec::43}\vspace{-10pt}
Table~\ref{Table::covariate_effects} summarizes the posterior distributions of the four covariate coefficients in the refit. Each coefficient is a probit-scale association in the sampled population, conditional on the tree ensemble, which includes the study indicator, and on the other covariates (Section~\ref{Sec::22}). The simulation studies in Section~\ref{Sec::3} show that spBART attenuates these coefficients and that strong gene--covariate dependence can bias them, so we interpret their magnitudes cautiously.

Race has the largest coefficient, with a posterior mean of $-0.59$ for European American (EA) relative to African American (AA) participants and a 95\% credible interval that includes zero. Because 88 of the 93 AA participants are MMEP cases, the data contain little information to separate race from study membership. We therefore retain race as an adjustment variable and do not interpret its coefficient as a racial difference in MM risk.

The BMI coefficient has a posterior mean of $-0.26$ and a posterior probability of 0.94 of being negative, and its credible interval also includes zero. Because BMI refers to the time of diagnosis in MMEP and is computed from current height and weight in BC case-control (Section~\ref{Sec::21}), this coefficient may also reflect the difference in measurement between the studies, including any weight loss after diagnosis among BC cases. Previous studies differ on this association, with a positive association in the meta-analysis of \citet{wallin2011body} and a nonsignificant one in \citet{blair2005anthropometric}, and our conditional estimate does not resolve the difference. We find little evidence of an association with sex. Age has a small positive coefficient of 0.01 per year, with a posterior probability of 0.96 of being positive. Because mean age and the proportion of cases both differ between the studies (Table~\ref{Table21}), we interpret this association only within the pooled sample.

The AMEs of age, sex and BMI in Table~\ref{Table::covariate_effects} are small, at most 5.5 percentage points in absolute value, so the variation in the predicted probabilities comes mainly from the genes and the study indicator in the tree ensemble. Older participants and participants with BMI below 25~kg/m$^2$ are somewhat more likely to be predicted as cases, in agreement with the signs of the coefficients, and sex makes little difference. Every AME interval includes zero, so these associations are suggestive only. The race AME of $-0.114$, like the race coefficient, mostly reflects study membership. In the simulations, the AME intervals of spBART stayed close to nominal coverage when its latent coefficients were attenuated (Section~\ref{Sec::35}), and we therefore regard the AMEs as the more reliable summary of the covariate associations. Because the AMEs are computed in the sampled case--control data, they describe differences in the predicted probability of being a case in this sample, not how much each covariate changes an individual's risk of MM.

\begin{table}[ht]
\centering
\caption{Posterior summaries of the covariate coefficients, on the probit scale, in the spBART refit to the development set. $P(\beta > 0)$ and $P(\beta < 0)$ are the posterior probabilities that the coefficient is positive and negative. AME indicates the average marginal effect on the probability scale, averaged over the 369 validation participants (for age, the effect of a 10-year increase). EA, European American; AA, African American.}
\label{Table::covariate_effects}
\resizebox{\textwidth}{!}{%
\begin{tabular}{lccccc}
\toprule
Variable & Posterior Mean & 95\% Credible Interval & $P(\beta > 0)$ & $P(\beta < 0)$ & AME (95\% Credible Interval) \\
\midrule
Race (EA vs.\ AA) & -0.59 & (-1.38, 0.10) & 0.05 & 0.95 & $-0.114$ ($-0.239$, $0.022$) \\
BMI ($\geq 25$ vs.\ $<25$) & -0.26 & (-0.58, 0.06) & 0.06 & 0.94 & $-0.055$ ($-0.123$, $0.013$) \\
Sex (Male vs.\ Female) & 0.07 & (-0.24, 0.40) & 0.67  & 0.33 & $0.015$ ($-0.052$, $0.084$) \\
Age & 0.01  & (0.00, 0.03)  & 0.96 & 0.04 & $0.026$ ($-0.002$, $0.051$) \\
\bottomrule
\end{tabular}}
\end{table}

\section{Simulation studies}\label{Sec::3}\vspace{-10pt}

\subsection{Overview}
We use simulation to examine what the semiparametric structure gains and loses in covariate inference, prediction and gene selection. We compare spBART with three methods. Dirichlet BART \citep{linero2018bayesian}, fitted with \texttt{pbart} in the \textsf{R} package \texttt{BART} \citep{sparapani2021nonparametric}, places the covariates in the tree ensemble with the genes, where they compete with the genes for splits. \texttt{stan4bart} \citep{dorie2022stan} has a linear term, updated by Hamiltonian Monte Carlo, but no sparsity prior on the splitting variables. The logistic elastic net \citep{zou2005regularization} leaves the coefficients of $\mathbf{Z}$ unpenalized. Our proposed spBART and Dirichlet BART use the same tree priors, number of trees, Dirichlet prior and MCMC settings. To compare spBART with Dirichlet BART under identical priors, we fit spBART in the simulations with two settings of \texttt{pbart} \citep{sparapani2021nonparametric}: the leaf prior $\sigma_\mu=3/(k\sqrt{T})$ with $k=2$, and a latent mean centered at the fixed offset $\Phi^{-1}(\bar Y)$, where $\bar Y$ is the proportion of cases in the training data. Differences between the two methods therefore reflect where the covariates enter the model, not the prior. The MM application uses $\sigma_\mu=0.5/\sqrt{T}$ and no offset (Section~\ref{Sec::223}).

Unless stated otherwise, we fit each method once to the whole simulated training set and evaluate prediction on an independent test set. The Bayesian methods select genes by the rule of Section~\ref{Sec::242} at $\alpha=0.05$, and the elastic net selects the genes with nonzero coefficients. This protocol omits the cross-validation union used in the application, which we evaluate separately in Section~\ref{Sec::36}. In the simulations, $\widetilde{\mathbf{X}}$ contains genes only. Most studies simulate outcomes from one of two data-generating models (DGMs) defined in Appendix~\ref{App::B}. DGM~1 is a sum of 20 shallow trees in which ten signal genes act through marginal effects and interactions with the covariates. DGM~2 combines 15 shallow trees in the same genes, including gene--gene interactions, with linear effects of age, sex, race and BMI.

We compare covariate associations on the probability scale by the AMEs in~\eqref{Eq::AME}. The latent predictor $\eta$ is the sum of the tree ensemble and the linear term for spBART and \texttt{stan4bart}, and the tree ensemble alone for Dirichlet BART, for which the AME is a partial-dependence contrast. We use a fixed subset $\mathcal{A}$ of 300 training subjects to limit computation and evaluate the true AME on the same subjects. We report the coverage of the 95\% credible intervals and the detection rate, defined by exclusion of zero from the interval. All covariate effects are nonzero except in Study~2, so detection measures power in the other studies, and Study~2 also reports false-positive rates.

Section~\ref{Sec::31} examines what the linear part gains for covariate inference (Studies~1 and~2), Section~\ref{Sec::35} examines gene selection and covariate inference in semi-synthetic data built on the observed 5hmC matrix (Study~3), and Section~\ref{Sec::36} evaluates the cross-validation union used in the application (Study~4). Appendix~\ref{App::B} gives all designs, including the data-generating trees, and the full results. It also reports four further studies: fitting without screening with up to 10{,}000 genes (Study~5), sensitivity to the prior variance of $\boldsymbol{\beta}$ and the contributions of the two components to prediction (Study~6), one long chain versus many short chains (Study~7), and smooth, correlated and partially misspecified models (Study~8). Study~5 shows that, with independent genes and at least 1{,}000 participants, spBART selects genes and predicts about as well as Dirichlet BART without screening, and Study~6 shows that its results vary little for $\sigma_\beta^2\geq1$, a range that includes our default of 10.

\vspace{-10pt}
\subsection{Covariate adjustment and inference}\label{Sec::31}\vspace{-10pt}

\paragraph{Study 1: a strong confounder.}
In this study, we investigate whether adjustment for a strong confounder prevents the selection of null genes that are correlated with it. We generate $n=1{,}500$ training subjects and an independent test set of 1{,}000, with $p\in\{2{,}000,5{,}000\}$ genes. Eight covariates enter the probit predictor linearly: age, sex, race, BMI, smoking, education, physical activity and family history. Age has a strong effect, $\beta_{\mathrm{age}}\in\{0.5,1.0,1.5\}$, and the other seven covariates have coefficients of 0.10--0.40 in absolute value. Ten signal genes contribute through the DGM~2 trees and are independent of the covariates. Thirty further genes have correlation $\rho\in\{0.5,0.8\}$ with age but are associated with the outcome only through age, so selecting any of them is a false discovery. Crossing the two values of $p$, the three values of $\beta_{\mathrm{age}}$ and the two values of $\rho$ gives $2\times3\times2=12$ settings, each with 100 replicates.

\begin{figure}[t]
\centering
\includegraphics[width=0.92\textwidth]{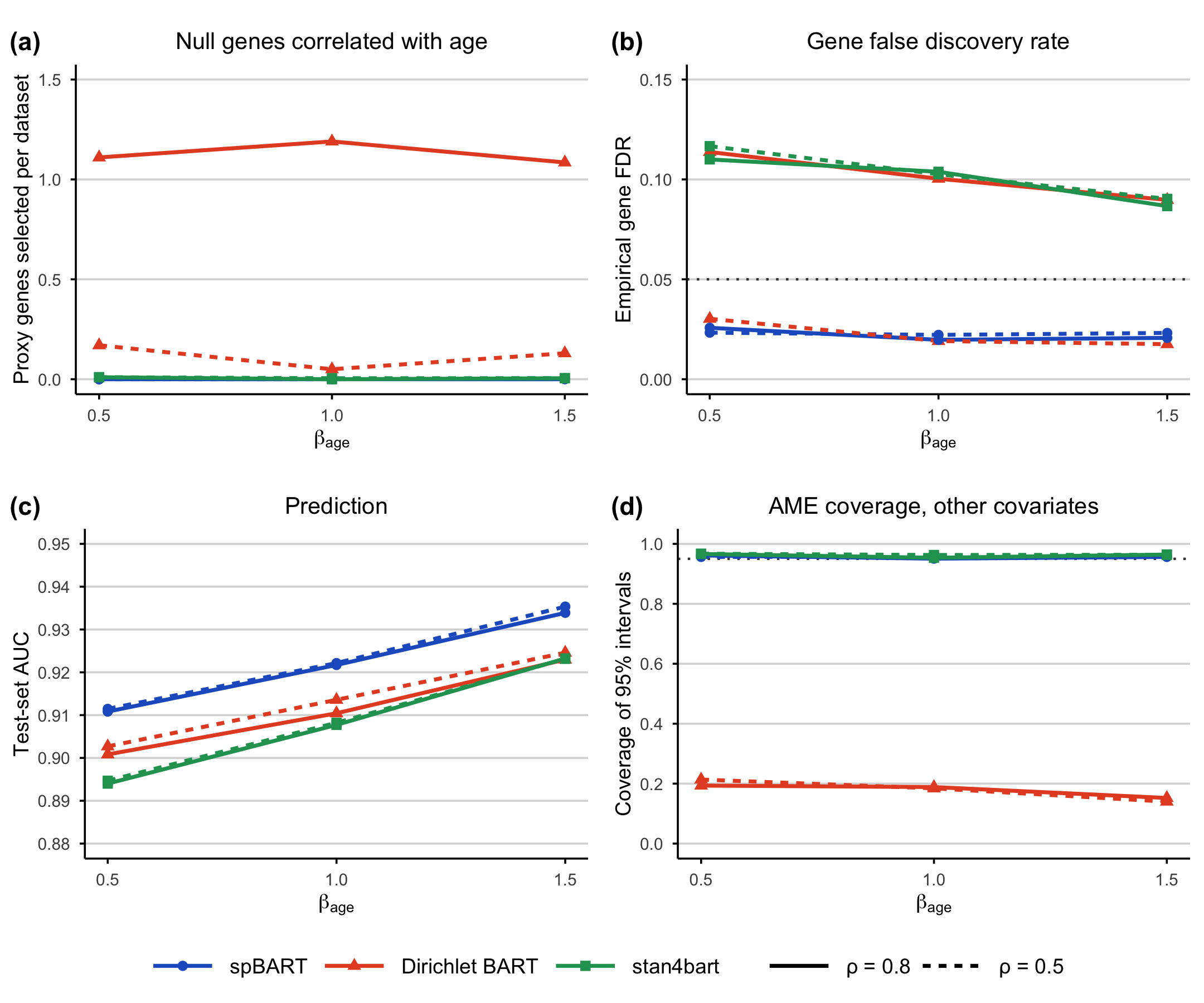}
\caption{(a) Number of selected null genes correlated with age, (b) empirical gene FDR, (c) test-set AUC and (d) coverage of 95\% intervals for the AMEs of the seven covariates other than age, as functions of the age coefficient, for a correlation between age and these null genes of $\rho=0.8$ (solid) or $\rho=0.5$ (dotted). Averages over 100 replicates and $p\in\{2{,}000,5{,}000\}$. }
\label{fig:sim1}
\end{figure}

With age in the linear part, spBART almost never selects the null genes correlated with age (Figure~\ref{fig:sim1}a). Dirichlet BART, in which age competes with the genes for splits, selects about one of them per dataset when $\rho=0.8$. Its empirical gene FDR is therefore 0.07--0.11 at $\rho=0.8$, whereas that of spBART is 0.01--0.03 in every setting. \texttt{stan4bart}, which has no sparsity prior on the splitting variables, has an empirical gene FDR of 0.06--0.15.

spBART also estimates the covariate effects more accurately. For the seven covariates other than age, the 95\% AME intervals of spBART and \texttt{stan4bart} cover the true effect in 95--97\% of datasets, whereas those of Dirichlet BART cover it in only 13--23\%, because its AMEs are shrunk toward zero and average 15\% of the true effects (Figure~\ref{fig:sim1}d). The test AUC of spBART is 0.006--0.016 above that of Dirichlet BART, with paired Monte Carlo standard errors below 0.003, and 0.011--0.018 above that of \texttt{stan4bart}.

These results assume that age acts linearly on the probit scale. In Study~8 (Appendix~\ref{App::sim8}), age and BMI act nonlinearly and interact with two genes. Placing them in the linear part then lowers the AUC of spBART by 0.007--0.009 relative to Dirichlet BART, and moving them into the trees restores a comparable AUC. The correlations between age and the genes in Study~1 are also stronger than those in the application, where the largest absolute age--gene correlation is 0.15 among the 115 screened genes and 0.22 among 2{,}000 randomly chosen features. Appendix~\ref{App::sim1} gives the full design and results.

\begin{figure}[t]
\centering
\includegraphics[width=0.92\textwidth]{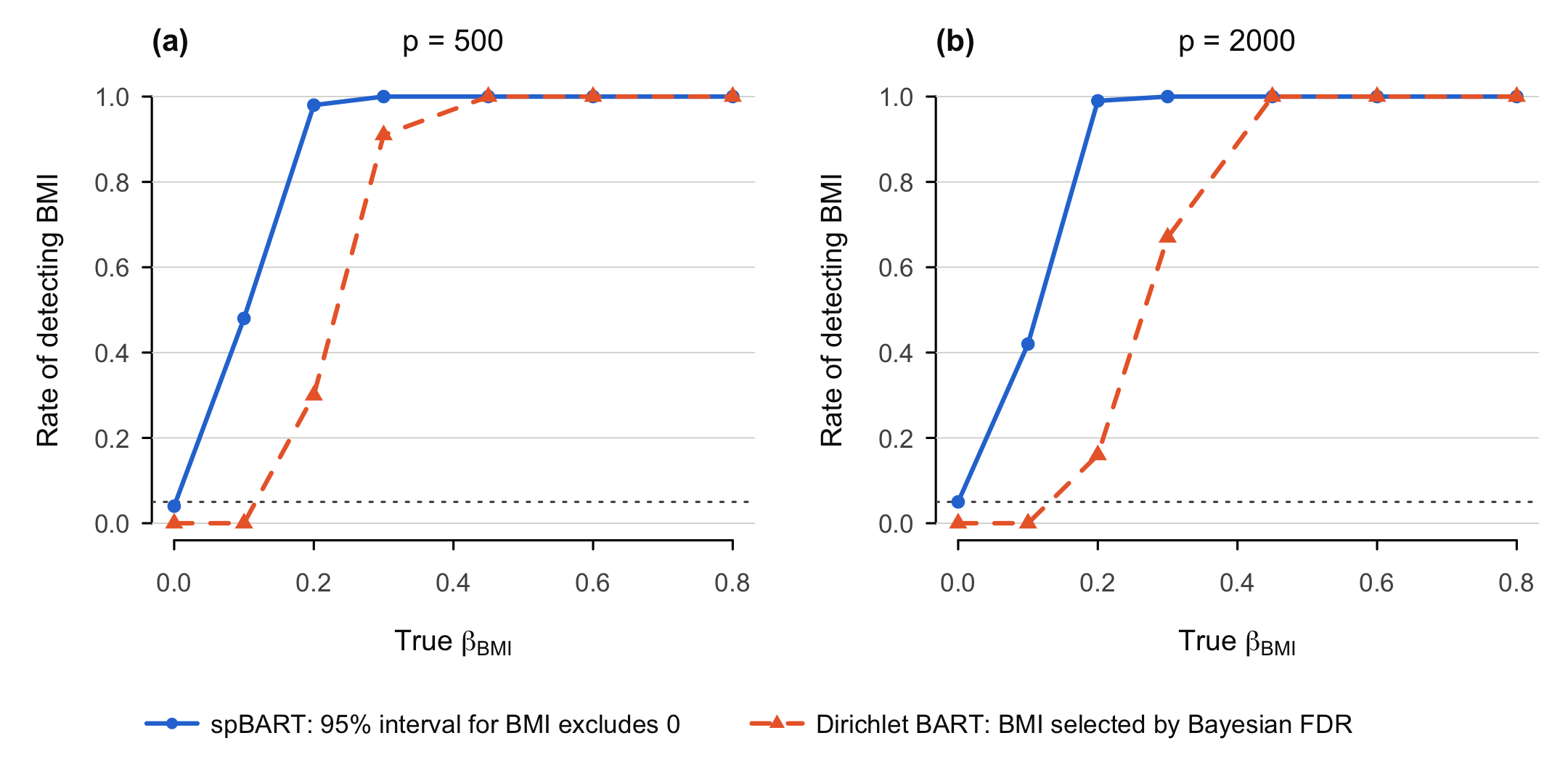}
\caption{BMI effect detection in Study~2 for (a) $p=500$ and
(b) $p=2{,}000$, based on 100 replicates per coefficient value.
spBART detects an effect when the 95\% credible interval for
$\beta_{\mathrm{BMI}}$ excludes zero. Dirichlet BART selects BMI using the
Bayesian FDR rule on covariate PIPs at $\alpha=0.05$.
Rates at $\beta_{\mathrm{BMI}}=0$ are false-positive rates.
The dotted line marks 0.05.}
\label{fig:sim2}
\end{figure}

\paragraph{Study 2: a small covariate effect.}
We examine how small a BMI association each method can detect. We use DGM~2 with $\boldsymbol{\beta}=(0.50,0.60,-0.40,\beta_{\mathrm{BMI}})$ for age, sex, race and BMI. We vary $\beta_{\mathrm{BMI}}\in\{0,0.1,0.2,0.3,0.45,0.6,0.8\}$, with $n=1{,}500$, $p\in\{500,2{,}000\}$ and 100 replicates per setting. spBART detects BMI when the 95\% credible interval of its coefficient excludes zero, and Dirichlet BART detects BMI when the Bayesian FDR rule, applied to the PIPs of the four covariates, selects it. The detection rate at $\beta_{\mathrm{BMI}}=0$ is the false-positive rate.

spBART detects smaller effects more often (Figure~\ref{fig:sim2}). At $\beta_{\mathrm{BMI}}=0.1$, the detection rates are 42--48\% for spBART and 0\% for Dirichlet BART, and at 0.2 they are 98--99\% and 16--30\%. Both methods reach 100\% at $\beta_{\mathrm{BMI}}\geq0.45$. The false-positive rates are 4--5\% for spBART and 0\% for Dirichlet BART. If we calibrate the PIP threshold of Dirichlet BART to a false-positive rate of 5\% using the simulated null distribution, which is unavailable in practice, its detection rate at $\beta_{\mathrm{BMI}}=0.2$ rises to 32--74\%, still below the 98--99\% of spBART (Appendix~\ref{App::sim2}).

Gene selection and prediction change little across the values of $\beta_{\mathrm{BMI}}$. Both methods have a gene sensitivity of at least 0.998. The empirical gene FDR is 0.013--0.025 for spBART and 0.008--0.012 for Dirichlet BART, and the test AUCs are 0.919--0.920 and 0.917--0.918, respectively. The latent-scale coefficient estimates of spBART are attenuated, however, and the coverage of the BMI interval falls from 95--96\% at $\beta_{\mathrm{BMI}}=0$ to 50--51\% at 0.8. The credible intervals thus detect small effects in this design, but the estimates of larger latent-scale effects are biased toward zero, and Study~6 (Appendix~\ref{App::sim6}) shows that this attenuation persists under diffuse priors on $\boldsymbol{\beta}$.

\vspace{-10pt}
\subsection{Synthetic data built on the observed 5hmC matrix}\label{Sec::35}\vspace{-10pt}

Study~3 keeps the dependence among the observed genes and covariates. We simulate outcomes for the development-set predictors ($n=500$) to train the models and for the validation-set predictors ($n=369$) to test them. The candidate genes are the 115 screened genes or a fixed draw of 2{,}000 random features. Ten signal genes are chosen at random or as one gene and its nine most correlated neighbors. We use either the DGM~2 trees with empirical-quantile cutpoints or a smooth nonlinear $f$, together with the linear covariate effects of DGM~2. Crossing the two candidate sets, the two signal layouts and the two forms of $f$ gives $2\times2\times2=8$ scenarios, each with 200 replicates. The study indicator is omitted, so the study does not assess the joint adjustment for race and study membership, which are nearly collinear in the application.

\begin{figure}[t]
\centering
\includegraphics[width=0.92\textwidth]{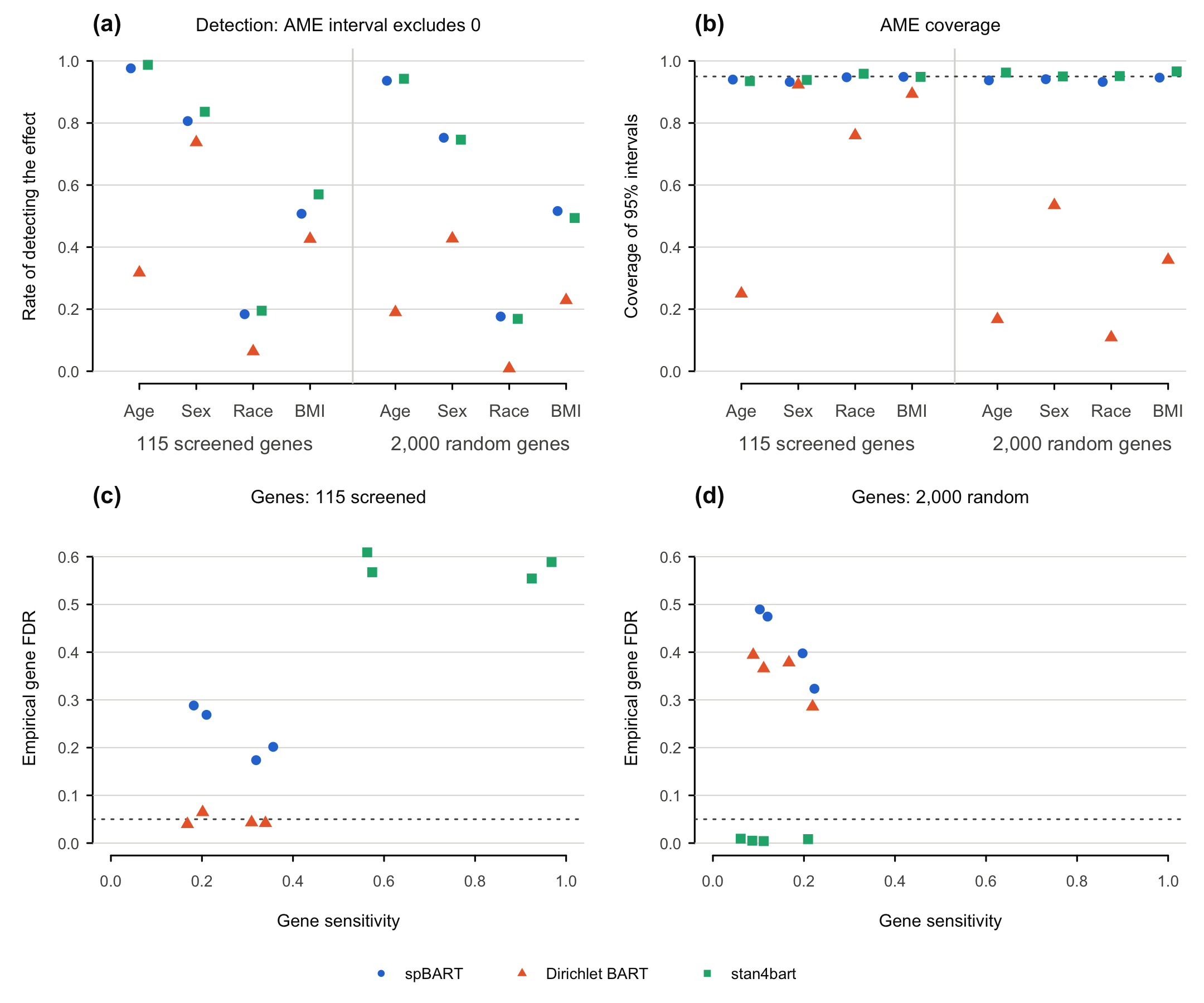}
\caption{Covariate inference and gene selection in Study~3 using
115 screened genes or 2,000 random features, with 200 replicates per
scenario. Panels show (a) detection by exclusion of zero from a
95\% AME interval, (b) AME interval coverage, (c) gene sensitivity,
and (d) empirical FDR. Covariate summaries are pooled over signal
layout and form of $f$, while gene results are shown by scenario. Dotted lines mark 0.95
coverage and the 0.05 FDR target.}
\label{fig:sim5}
\end{figure}

With the real predictors, the AME intervals of spBART keep close to nominal coverage (Figure~\ref{fig:sim5}). Pooled over the four scenarios of each gene set, the coverage of the 95\% AME intervals is 93--95\% for spBART, 94--97\% for \texttt{stan4bart} and 11--92\% for Dirichlet BART. spBART also detects the covariates by their AMEs more often than Dirichlet BART, for example in 94--98\% versus 19--32\% of datasets for age, and about as often as \texttt{stan4bart}. With the screened genes, the PIP rule of Dirichlet BART selects sex, race and BMI more often than the intervals of spBART detect them, yet its AME-based detection is lower. A covariate used in many splits may therefore still have an AME indistinguishable from zero. The latent coefficients of spBART remain attenuated, with posterior means of 0.42--0.80 of their true values.

Gene selection is much harder with the real predictors. Sensitivity is 10--36\% for spBART and 9--34\% for Dirichlet BART. With the screened genes, the empirical FDR is 0.17--0.29 for spBART and 0.04--0.06 for Dirichlet BART, rising to 0.32--0.49 and 0.29--0.39 with 2{,}000 genes. Both methods use the same selection rule, and the source of the higher FDR of spBART remains unresolved. The median absolute correlation with the covariates is 0.12 among its false discoveries and 0.10 among the other null genes, which suggests that correlation with the covariates is not the main source. Knockoffs have similar sensitivity and a lower empirical FDR of 0.11--0.32, which still exceeds their 0.05 target. \texttt{stan4bart} has the highest AUC, 0.02--0.06 above that of spBART, and the least biased coefficients, but its gene selection varies sharply: it selects many signal and null genes from the screened set and few genes from the larger set. In an application-sized scenario of Study~7 (Appendix~\ref{App::sim7}), selecting genes from the pooled draws of 10 or 50 chains reduces the empirical FDR of spBART from 0.19 to 0.01--0.02, while its sensitivity falls from 0.37 to 0.25--0.26.

\vspace{-10pt}
\subsection{Selection with the cross-validation union}\label{Sec::36}\vspace{-10pt}

Study~4 (Appendix~\ref{App::sim4}) evaluates the selection procedure of the application, in which we apply the Bayesian FDR rule within five cross-validation folds, take the union of the selected genes and refit the model to them. Both spBART and Dirichlet BART recover all ten signal genes in every setting. The union also admits null genes, with empirical FDRs of 0.06--0.16 for spBART and 0.02--0.08 for Dirichlet BART, which is consistent with our treatment of the union in the application as a set of candidate genes (Section~\ref{Sec::242}). The test AUCs are comparable when the covariates act linearly, and the AUC of spBART is 0.012--0.013 lower when the covariates act only through gene--covariate interactions, which its linear part cannot represent.

\vspace{-10pt}
\subsection{Summary}\label{Sec::37}\vspace{-10pt}
The above four simulation studies bear on the application in three respects. When the covariates act approximately linearly, placing them in the linear part gives AME intervals with close to nominal coverage and detects small covariate effects that Dirichlet BART shrinks toward zero. The latent coefficients of spBART are attenuated, however, so the covariate coefficients in Table~\ref{Table::covariate_effects} are more informative through their signs and posterior probabilities than through their magnitudes. Gene selection is much harder with the observed 5hmC matrix and 500 participants than with independent genes. spBART then recovers 10--36\% of the signal genes, and 17--29\% of the genes it selects from the screened set are null. The union of fold-specific selections recovers the signal genes but also admits null genes. These results support treating the eight genes in~\eqref{Eq::selected_genes} as candidates and assessing each by the stability of its selection across folds, chains and the BC-only analysis (Section~\ref{Sec::44}).

\section{Discussion}\label{Sec::5}\vspace{-15pt}

In the pooled case--control data, spBART selects eight candidate genes whose circulating 5hmC levels are associated with MM status, and the model discriminates cases from controls within BC case-control, where study membership is constant, with a validation AUC of 0.92. The support differs among the eight genes. \textit{IL1RAP} is selected in every fold of every analysis, including the analysis of BC case-control alone, and \textit{ST5} enters every four-chain union and two of the five BC-only folds. The other six genes are selected in few fits once we change the Markov chain or the data. Previous work on IL-1/IL-6 signaling in MM gives a biological reason to study \textit{IL1RAP} further \citep{donovan1998contrast,lust2009induction}. Because cfDNA comes from several cell types, these associations may reflect myeloma cells, the host response or both, and our data cannot identify the source of the 5hmC differences.

The simulations favor the semiparametric structure when the covariate associations are approximately linear on the probit scale, and Dirichlet BART predicts better when the covariates act only through interactions with genes. In the semi-synthetic study built on the observed 5hmC matrix, spBART selects more null genes than Dirichlet BART, although both use the same selection rule, and we have not identified the reason.

Our selection procedure takes the union of five fold-specific selections, which favors sensitivity. The posterior expected FDR bound applies within each fold-specific fit, and Study~4 shows that the union admits additional null genes. The refit cannot rank the candidates either, because each has a PIP of 1.00 in a model restricted to them. We therefore measure the support for each candidate by its selection across folds, chains and studies. In our data this support varies with the Markov chain: when the draws of four chains are pooled, five of the eight genes drop out of the union, and \textit{NF2}, which no fold of the single-chain analysis selects, enters every chain's union. Study~7 indicates that selecting from the pooled draws of several long chains reduces false discoveries in data of the size of our application, at some cost in sensitivity.

The main limitation of the data is the design. All controls come from BC case-control and MMEP contributes only cases, so study membership alone gives a validation AUC of 0.77. The analyses within BC case-control show discrimination that study membership cannot explain, but no analysis of these data can evaluate the model among MMEP participants, who include 88 of the 93 AA participants. Race is nearly confounded with study membership, and BMI is defined differently in the two studies, so we use the race coefficient only for adjustment and interpret the BMI coefficient with caution. The covariate coefficients also describe the sampled population, because the probit model lacks the invariance of the logistic model under case--control sampling \citep{prentice1979logistic}, and the simulations show that spBART attenuates them while the probability-scale AMEs are less affected. Extending spBART to a logistic link would give the covariate coefficients the interpretation of conditional odds ratios that carries over from the case--control sample to the population.

The next step is independent replication of \textit{IL1RAP} and \textit{ST5}, including the direction of their 5hmC differences, in a study whose cases and controls come from the same population and include enough AA participants. Matched tissue or cell-type-specific profiles would help to distinguish changes in myeloma cells from those of the host response.

\vspace{-15pt}
\section*{Code/software availability}\vspace{-15pt}
The code implementing the proposed spBART methodology and reproducing the analyses in this manuscript is publicly available at \url{https://github.com/SBstats/spBART-MM-case-control}.

\vspace{-15pt}
\section*{Acknowledgements}\vspace{-15pt}
This research was supported, in part, by the National Institutes of Health grants: R01CA223662, R33CA269100, R56CA282891, R01CA280637, and R21CA283588.

\bibliographystyle{apalike}
\bibliography{references}

\newpage

\appendix

\section{Posterior computation}\label{App::A}

In this section, we describe the Markov chain Monte Carlo (MCMC) algorithm for posterior inference in the spBART model. 

\subsection{Latent variable augmentation}

Following \citet{albert1993bayesian}, we introduce continuous latent variables $U_i$ for $i = 1, \ldots, N$ such that the observed binary outcome satisfies $Y_i = \mathbb{I}(U_i > 0)$. Under the spBART model, the latent variables follow
\[
U_i \mid f, \boldsymbol{\beta} \sim \mathcal{N}\big(f(\widetilde{\mathbf{X}}_i) + \mathbf{Z}_i^\top \boldsymbol{\beta},\, 1\big),
\]
where $f(\widetilde{\mathbf{X}}_i) = \sum_{t=1}^T g(\widetilde{\mathbf{X}}_i; \mathcal{T}_t, \boldsymbol{\mu}_t)$ denotes the BART ensemble evaluated at the augmented 5hmC signatures predictor vector $\widetilde{\mathbf{X}}_i$, and $\mathbf{Z}_i^\top \boldsymbol{\beta}$ is the linear contribution from covariates.

Conditional on the current values of $f$ and $\boldsymbol{\beta}$, each latent variable $U_i$ is sampled independently from a truncated normal distribution:
\[
U_i \mid Y_i, f, \boldsymbol{\beta} \sim
\begin{cases}
\mathcal{TN}_{(0, \infty)}\big(\mu_i, 1\big), & \text{if } Y_i = 1, \\[4pt]
\mathcal{TN}_{(-\infty, 0]}\big(\mu_i, 1\big), & \text{if } Y_i = 0,
\end{cases}
\]
where $\mu_i = f(\widetilde{\mathbf{X}}_i) + \mathbf{Z}_i^\top \boldsymbol{\beta}$ and $\mathcal{TN}_{(a,b)}(\mu, \sigma^2)$ denotes the normal distribution with mean $\mu$ and variance $\sigma^2$ truncated to the interval $(a, b)$.

\subsection{Gibbs update for regression coefficients}

Given the latent variables $\mathbf{U} = (U_1, \ldots, U_N)^\top$ and the tree ensemble $f$, define the partial residuals $\mathbf{R} = \mathbf{U} - \mathbf{f}$, where $\mathbf{f} = (f(\widetilde{\mathbf{X}}_1), \ldots, f(\widetilde{\mathbf{X}}_N))^\top$. The model for these residuals is a standard linear regression:
\[
\mathbf{R} = \mathbf{Z} \boldsymbol{\beta} + \boldsymbol{\varepsilon}, \qquad \boldsymbol{\varepsilon} \sim \mathcal{N}(\mathbf{0}, I_N),
\]
where $\mathbf{Z}$ is the $N \times J$ design matrix with rows $\mathbf{Z}_i^\top$. Under the conjugate prior $\boldsymbol{\beta} \sim \mathcal{N}(\mathbf{0}, \sigma_\beta^2 I_J)$, the full conditional distribution is multivariate normal:
\[
\boldsymbol{\beta} \mid \mathbf{U}, f \sim \mathcal{N}\big(\mathbf{V}_\beta \mathbf{Z}^\top \mathbf{R},\, \mathbf{V}_\beta\big),
\]
where $\mathbf{V}_\beta = (\mathbf{Z}^\top \mathbf{Z} + \sigma_\beta^{-2} I_J)^{-1}$. In our implementation, we fix $\sigma_\beta^2 = 10$, corresponding to a weakly informative prior. If a hierarchical prior is used for $\sigma_\beta$, it is updated via a Metropolis--Hastings or Gibbs step as applicable.

\subsection{Updating the tree ensemble}

Conditional on $\boldsymbol{\beta}$ and $\mathbf{U}$, the tree ensemble $f$ is updated using the Bayesian backfitting algorithm \citep{chipman2010bart}. Define the partial residuals for tree $t$ as
\[
R_{it} = U_i - \mathbf{Z}_i^\top \boldsymbol{\beta} - \sum_{s \neq t} g(\widetilde{\mathbf{X}}_i; \mathcal{T}_s, \boldsymbol{\mu}_s).
\]
Each tree $(\mathcal{T}_t, \boldsymbol{\mu}_t)$ is then updated in turn by treating $R_{it}$ as the response in a Gaussian regression with unit variance. Tree structure updates (grow, prune, change, swap) are proposed via Metropolis--Hastings steps with acceptance probabilities determined by the product of the likelihood ratio and the prior ratio under the regularization prior on tree depth. Terminal node parameters $\boldsymbol{\mu}_t$ are sampled from their conjugate normal full conditionals.

Under the Dirichlet BART extension \citep{linero2018bayesian}, the probability of selecting variable $j$ for a split is governed by the splitting probability $\pi_j$ rather than being uniform. 

\subsection{Updating splitting probabilities}

The Dirichlet BART prior places a symmetric Dirichlet distribution on the splitting probabilities $(\pi_1, \ldots, \pi_{D^*})$, where $D^* = D + 1$ is the number of candidate splitting variables (screened 5hmC signatures plus the study indicator). Let $s_j$ denote the total number of splits on variable $j$ across all $T$ trees in the current ensemble. By conjugacy, the full conditional distribution is
\[
(\pi_1, \ldots, \pi_{D^*}) \mid \{s_j\} \sim \mathrm{Dirichlet}\Big(\tfrac{\xi}{D^*} + s_1,\, \ldots,\, \tfrac{\xi}{D^*} + s_{D^*}\Big),
\]
where $\xi > 0$ is the concentration parameter. Variables that are frequently used for splitting receive larger posterior splitting probabilities, reinforcing their selection in subsequent iterations.

\section{Details on simulation studies}
\label{App::B}

We first describe the shared data-generating models (DGMs), predictor generation, and performance measures, then report each study.
Studies~1--4 supplement Section~\ref{Sec::3} with design details and
result tables. Studies~5--8 are reported here in full. In the tables,
sp, B, s4b, EN, and KO denote spBART, BART (\texttt{pbart} with the
Dirichlet prior), \texttt{stan4bart}, elastic net, and model-X
knockoffs, respectively.

\paragraph{Genes, covariates, and outcomes.}
Unless stated otherwise, genes are generated independently as
$X_{ij}\stackrel{\mathrm{iid}}{\sim}\mathcal{N}(0,1)$, with the first
$s_0=10$ serving as signals. Covariates are generated independently
of the genes and one another: standardized age
$Z_1\sim\mathcal{N}(0,1)$, sex
$Z_2\sim\mathrm{Bernoulli}(0.5)$, race
$Z_3\sim\mathrm{Bernoulli}(0.3)$, and standardized BMI
$Z_4\sim\mathcal{N}(0,1)$. Outcomes follow
\[
Y_i\sim\mathrm{Bernoulli}\{\Phi(f(\mathbf{X}_i,\mathbf{Z}_i))\},
\]
where $f$ is a DGM defined below and $\Phi$ is the standard normal
CDF. Each replicate includes an independent test set of 1,000
subjects generated in the same way. Semi-synthetic designs instead
use the real validation-set predictors with simulated outcomes.

\paragraph{Data-generating trees.} DGM~1 and DGM~2 are constructed as sums of shallow regression trees. Two considerations guide the design of these data-generating trees:
\begin{enumerate}
    \item \textbf{Tree complexity:} We restrict trees to depth 1 (marginal effects) and depth 2 (two-way interactions). The default BART priors favor shallow trees through the regularization prior on tree depth, so deeper, more complex trees would be poorly approximated by the fitted ensembles. The goal of this simulation study is to compare the two competing methods when the true data-generating process is known, not to evaluate BART's ability to recover arbitrarily complex tree structures.
    \item \textbf{Signal magnitude:} Terminal node values are chosen to avoid extreme values of the linear predictor inside the probit link. If the sum of tree contributions becomes too large in magnitude, the response probabilities $\Phi(f(\mathbf{X}, \mathbf{Z}))$ approach 0 or 1, resulting in near-deterministic outcomes with minimal variation. We calibrate the terminal values (ranging from $\pm 0.35$ to $\pm 0.40$) so that the aggregated signal produces balanced case--control proportions.
\end{enumerate}

\subsubsection*{Cutpoint specification}

Tree cutpoints are defined as fixed population quantiles from the $\mathcal{N}(0,1)$ distribution to ensure the true regression function is fixed across simulation replications. Let $\Phi^{-1}(\cdot)$ denote the standard normal quantile function.

The cutpoints for the 10 signal genes are:
\begin{align*}
c_1 &= \Phi^{-1}(0.45) = -0.126, & c_2 &= \Phi^{-1}(0.55) = 0.126, & c_3 &= \Phi^{-1}(0.40) = -0.253, \\
c_4 &= \Phi^{-1}(0.60) = 0.253, & c_5 &= \Phi^{-1}(0.50) = 0, & c_6 &= \Phi^{-1}(0.50) = 0, \\
c_7 &= \Phi^{-1}(0.35) = -0.385, & c_8 &= \Phi^{-1}(0.65) = 0.385, & c_9 &= \Phi^{-1}(0.42) = -0.202, \\
c_{10} &= \Phi^{-1}(0.58) = 0.202.
\end{align*}

The cutpoints for continuous covariates are:
\[
c_{\text{age}} = \Phi^{-1}(0.55) = 0.126, \qquad c_{\text{BMI}} = \Phi^{-1}(0.40) = -0.253.
\]

\subsubsection*{DGM 1: Pure sum of trees}

DGM 1 consists of 20 regression trees with no linear covariate component.

\paragraph{Depth-1 trees: Marginal gene effects (Trees 1--10).}
Each signal gene $j \in \{1, \ldots, 10\}$ has exactly one marginal tree with terminal values $\pm 0.35$:
\begin{align*}
g_j(X_j) &= -0.35 \cdot \mathbb{I}(X_j \leq c_j) + 0.35 \cdot \mathbb{I}(X_j > c_j), \quad j = 1, \ldots, 10.
\end{align*}

\paragraph{Depth-2 trees: Gene--covariate interactions (Trees 11--20).}
Each signal gene has exactly one interaction tree with symmetric terminal values $\pm 0.40$:
\begin{align*}
g_{11}(X_1, Z_1) &= 0.40 \cdot \mathbb{I}(X_1 > c_1, Z_1 > c_{\text{age}}) - 0.40 \cdot \mathbb{I}(X_1 \leq c_1, Z_1 \leq c_{\text{age}}) \\
g_{12}(X_2, Z_2) &= 0.40 \cdot \mathbb{I}(X_2 > c_2, Z_2 = 1) - 0.40 \cdot \mathbb{I}(X_2 \leq c_2, Z_2 = 0) \\
g_{13}(X_3, Z_4) &= 0.40 \cdot \mathbb{I}(X_3 > c_3, Z_4 > c_{\text{BMI}}) - 0.40 \cdot \mathbb{I}(X_3 \leq c_3, Z_4 \leq c_{\text{BMI}}) \\
g_{14}(X_4, Z_3) &= 0.40 \cdot \mathbb{I}(X_4 > c_4, Z_3 = 1) - 0.40 \cdot \mathbb{I}(X_4 \leq c_4, Z_3 = 0) \\
g_{15}(X_5, Z_1) &= 0.40 \cdot \mathbb{I}(X_5 > c_5, Z_1 > c_{\text{age}}) - 0.40 \cdot \mathbb{I}(X_5 \leq c_5, Z_1 \leq c_{\text{age}}) \\
g_{16}(X_6, Z_2) &= 0.40 \cdot \mathbb{I}(X_6 > c_6, Z_2 = 1) - 0.40 \cdot \mathbb{I}(X_6 \leq c_6, Z_2 = 0) \\
g_{17}(X_7, Z_4) &= 0.40 \cdot \mathbb{I}(X_7 > c_7, Z_4 > c_{\text{BMI}}) - 0.40 \cdot \mathbb{I}(X_7 \leq c_7, Z_4 \leq c_{\text{BMI}}) \\
g_{18}(X_8, Z_3) &= 0.40 \cdot \mathbb{I}(X_8 > c_8, Z_3 = 1) - 0.40 \cdot \mathbb{I}(X_8 \leq c_8, Z_3 = 0) \\
g_{19}(X_9, Z_1) &= 0.40 \cdot \mathbb{I}(X_9 > c_9, Z_1 > c_{\text{age}}) - 0.40 \cdot \mathbb{I}(X_9 \leq c_9, Z_1 \leq c_{\text{age}}) \\
g_{20}(X_{10}, Z_2) &= 0.40 \cdot \mathbb{I}(X_{10} > c_{10}, Z_2 = 1) - 0.40 \cdot \mathbb{I}(X_{10} \leq c_{10}, Z_2 = 0)
\end{align*}
where $Z_1$ denotes age, $Z_2$ denotes sex, $Z_3$ denotes race, and $Z_4$ denotes BMI.

\subsubsection*{DGM 2: Gene-only trees with linear covariate effects}

DGM 2 consists of 15 gene-only regression trees plus linear covariate effects.

\paragraph{Depth-1 trees: Marginal gene effects (Trees 1--10).}
Each signal gene $j \in \{1, \ldots, 10\}$ has exactly one marginal tree with terminal values $\pm 0.40$:
\begin{align*}
g_j(X_j) &= -0.40 \cdot \mathbb{I}(X_j \leq c_j) + 0.40 \cdot \mathbb{I}(X_j > c_j), \quad j = 1, \ldots, 10.
\end{align*}

\paragraph{Depth-2 trees: Gene--gene interactions (Trees 11--15).}
Five interaction trees pair genes with symmetric terminal values $\pm 0.35$:
\begin{align*}
g_{11}(X_1, X_2) &= 0.35 \cdot \mathbb{I}(X_1 > c_1, X_2 > c_2) - 0.35 \cdot \mathbb{I}(X_1 \leq c_1, X_2 \leq c_2) \\
g_{12}(X_3, X_4) &= 0.35 \cdot \mathbb{I}(X_3 > c_3, X_4 > c_4) - 0.35 \cdot \mathbb{I}(X_3 \leq c_3, X_4 \leq c_4) \\
g_{13}(X_5, X_6) &= 0.35 \cdot \mathbb{I}(X_5 > c_5, X_6 > c_6) - 0.35 \cdot \mathbb{I}(X_5 \leq c_5, X_6 \leq c_6) \\
g_{14}(X_7, X_8) &= 0.35 \cdot \mathbb{I}(X_7 > c_7, X_8 > c_8) - 0.35 \cdot \mathbb{I}(X_7 \leq c_7, X_8 \leq c_8) \\
g_{15}(X_9, X_{10}) &= 0.35 \cdot \mathbb{I}(X_9 > c_9, X_{10} > c_{10}) - 0.35 \cdot \mathbb{I}(X_9 \leq c_9, X_{10} \leq c_{10})
\end{align*}

\paragraph{Linear covariate effects.}
The linear component is given by
\[
\mathbf{Z}^\top \boldsymbol{\beta} = 0.50 \cdot Z_1 + 0.60 \cdot Z_2 - 0.40 \cdot Z_3 + 0.45 \cdot Z_4,
\]
where $Z_1$ (scaled age), $Z_2$ (sex), $Z_3$ (race), and $Z_4$ (scaled BMI) are defined at the beginning of this appendix.

\subsubsection*{DGM 3: smooth nonlinear gene effects}
To represent a regression function without true cutpoints, DGM~3 replaces the gene trees of DGM~2 by a smooth function of the ten signal genes,
\begin{align*}
f_3(\mathbf{X})=s\Big[&\sin\!\big(\tfrac{\pi}{2}X_1X_2\big)+0.5\,(X_3^2-1)+0.6\,X_4+\tanh(1.5\,X_5)+0.5\,X_6X_7\\
&+\big(e^{-X_8^2}-1/\sqrt{3}\big)+0.5\big(|X_9|-\sqrt{2/\pi}\big)+0.4\,X_{10}\Big],
\end{align*}
with $s=1.162$ chosen so that the standard deviation of $f_3$ equals that of the DGM~2 gene trees (1.786). The covariates enter linearly as in DGM~2.

\subsubsection*{Methods and settings}
\begin{itemize}
\item \textbf{spBART} is fitted as in Section~\ref{Sec::22} with $T=200$ trees, 2{,}000 burn-in iterations and 5{,}000 further iterations thinned by 5 ($Q=1{,}000$ draws), and $\sigma_\beta^2=10$. The leaf prior is $\mu_\ell\sim\mathcal{N}\{0,(3/(k\sqrt{T}))^2\}$ with $k=2$, each predictor has 100 candidate cutpoints, the Dirichlet prior uses $a=0.5$, $b=1$ and $\rho$ equal to the number of candidate splitting variables and is switched on halfway through burn-in, and, as in \texttt{pbart}, the latent mean includes the fixed offset $\Phi^{-1}(\bar Y)$, which centers $f$ at the sample prevalence. This leaf prior is the \texttt{pbart} default for binary outcomes, so spBART and BART share it.
\item \textbf{BART} is \texttt{pbart} \citep{sparapani2021nonparametric} with \texttt{sparse = TRUE}, applied to $(\mathbf{X},\mathbf{Z})$, with the same number of trees, leaf prior, cutpoints, Dirichlet hyperparameters (with $\rho$ equal to the number of genes plus covariates) and MCMC length. Gene PIPs are computed exactly as for spBART. The covariate PIPs are excluded from the gene FDR rule. To "select" a covariate with BART we apply the same Bayesian FDR rule at $\alpha=0.05$ to the covariate PIPs.
\item \textbf{\texttt{stan4bart}} \citep{dorie2022stan} is fitted with the probit link, $\mathbf{Z}$ as fixed effects and the genes in the BART term (200 trees), with four chains of 1{,}500 iterations of which 1{,}000 are warm-up. PIPs are computed from its split counts.
\item \textbf{Elastic net}: logistic regression with mixing parameter 0.5, $\mathbf{Z}$ unpenalized, penalty chosen by 10-fold cross-validation (\texttt{lambda.1se}); genes with nonzero coefficients are selected.
\item \textbf{Knockoffs} (Study~3): second-order Gaussian model-X knockoffs \citep{candes2018panning}, a lasso fitted to the genes, their knockoffs and $\mathbf{Z}$ (unpenalized), the coefficient-difference statistic and the knockoff threshold at target FDR 0.05 without the $+1$ offset of knockoff$+$ (which controls a modified FDR). With non-Gaussian real features the second-order construction is only approximate, so FDR control is not guaranteed in Study~3.
\end{itemize}

\subsubsection*{Performance measures}
Sensitivity is the proportion of the ten signal genes selected.
Empirical FDR is the proportion of selected genes that are null,
set to zero when none are selected. Both are averaged over replicates.
The test AUC and Brier score use posterior mean probabilities.
Coverage and detection are the proportions of replicates whose
95\% credible intervals contain the true coefficient or AME and
exclude zero, respectively. The true AME applies the contrast in
\eqref{Eq::AME} to the true regression function on the same 300 subjects.


\subsection{Study 1: adjustment for a strong confounder}\label{App::sim1}

 Age, BMI, education, and physical activity are generated from standard normal distributions. Sex, race, smoking, and family history are Bernoulli with probabilities
0.5, 0.3, 0.25, and 0.1, respectively. We use $n=1{,}500$ and
$p\in\{2{,}000,5{,}000\}$. The probit predictor combines the DGM~2
trees in genes 1--10 with the covariate term
\[
\begin{aligned}
&\beta_{\mathrm{age}}\,\mathrm{age}_i
+0.30\,\mathrm{sex}_i-0.30\,\mathrm{race}_i
+0.25\,\mathrm{BMI}_i\\
&\quad+0.20\,\mathrm{smoking}_i-0.15\,\mathrm{education}_i
-0.10\,\mathrm{activity}_i+0.40\,\mathrm{family\,history}_i.
\end{aligned}
\]
Genes 11--40 are null genes correlated with age,
$X_{ij}=\rho\,\mathrm{age}_i+\sqrt{1-\rho^2}\,\varepsilon_{ij}$,
with no outcome contribution conditional on age.
spBART and \texttt{stan4bart} include all eight covariates linearly,
while BART includes them in the trees.

Tables~\ref{tab:A1sel}--\ref{tab:A1cov} give scenario-specific results.
At $p=5{,}000$, sensitivity is 0.93--0.96 for spBART and 0.89--0.96
for BART, whose sensitivity declines more as the age effect increases.
spBART's coefficient means average approximately 0.85 of the true
values, with age coverage of 40\% pooled over the three age effects.
\texttt{stan4bart}'s coefficients are less biased, generally about
10\% above the truth. Both have near-nominal AME coverage
(Table~\ref{tab:A1inf}).

BART selects the seven smaller covariates in 0--29\% of datasets
(Table~\ref{tab:A1cov}). Its AME posterior means average 15\% of
the true effects, with shrinkage toward zero in 94\% of
covariate--replicate pairs. After multiplying bias by the sign of
the true effect, mean bias is $-3.4$ percentage points, compared
with 0.0 for spBART. spBART's AME means average 100\% of the true
effects. Mean interval widths are 1.5 percentage points for BART
and 6.5 for spBART.

\begin{table}[htbp]\centering
\caption{Study 1: gene selection. Sensitivity, empirical FDR, and the number of selected null genes among the 30 correlated with age, averaged over 100 replicates.}
\label{tab:A1sel}
\resizebox{0.85\textwidth}{!}{
\begin{tabular}{rlr ccc cccc ccc}
\toprule
& & & \multicolumn{3}{c}{Sensitivity} & \multicolumn{4}{c}{FDR} & \multicolumn{3}{c}{\# age-correlated nulls selected} \\
\cmidrule(lr){4-6}\cmidrule(lr){7-10}\cmidrule(lr){11-13}
$p$ & $\beta_{\rm age}$ & $\rho$ & sp & B & s4b & sp & B & s4b & EN & sp & B & s4b \\
\midrule
2000 & 0.5 & 0.5 & 1.00 & 1.00 & 1.00 & 0.02 & 0.03 & 0.15 & 0.63 & 0.00 & 0.16 & 0.02 \\
 & 0.5 & 0.8 & 1.00 & 1.00 & 1.00 & 0.03 & 0.11 & 0.13 & 0.64 & 0.00 & 1.17 & 0.02 \\
 & 1.0 & 0.5 & 1.00 & 1.00 & 1.00 & 0.02 & 0.01 & 0.12 & 0.61 & 0.00 & 0.00 & 0.01 \\
 & 1.0 & 0.8 & 1.00 & 1.00 & 1.00 & 0.01 & 0.09 & 0.13 & 0.59 & 0.00 & 1.12 & 0.00 \\
 & 1.5 & 0.5 & 1.00 & 1.00 & 1.00 & 0.01 & 0.01 & 0.12 & 0.56 & 0.01 & 0.00 & 0.01 \\
 & 1.5 & 0.8 & 1.00 & 1.00 & 1.00 & 0.01 & 0.07 & 0.12 & 0.55 & 0.00 & 0.86 & 0.01 \\
\addlinespace[2pt]
5000 & 0.5 & 0.5 & 0.96 & 0.96 & 0.98 & 0.03 & 0.03 & 0.09 & 0.69 & 0.00 & 0.18 & 0.00 \\
 & 0.5 & 0.8 & 0.96 & 0.96 & 1.00 & 0.02 & 0.11 & 0.09 & 0.65 & 0.00 & 1.05 & 0.00 \\
 & 1.0 & 0.5 & 0.95 & 0.93 & 0.99 & 0.03 & 0.03 & 0.08 & 0.64 & 0.00 & 0.10 & 0.00 \\
 & 1.0 & 0.8 & 0.95 & 0.94 & 0.98 & 0.03 & 0.11 & 0.07 & 0.67 & 0.00 & 1.26 & 0.00 \\
 & 1.5 & 0.5 & 0.94 & 0.89 & 0.98 & 0.03 & 0.03 & 0.06 & 0.57 & 0.00 & 0.26 & 0.00 \\
 & 1.5 & 0.8 & 0.93 & 0.91 & 0.98 & 0.03 & 0.11 & 0.06 & 0.58 & 0.00 & 1.31 & 0.00 \\
\bottomrule
\end{tabular}
}
\end{table}
\begin{table}[htbp]\centering
\caption{Study 1: coverage of 95\% AME intervals for age and averaged over the seven other covariates, mean absolute AME bias for those seven covariates in percentage points (pp), and test-set AUC.}
\label{tab:A1inf}
\resizebox{\textwidth}{!}{
\begin{tabular}{rlr ccc ccc ccc cccc}
\toprule
& & & \multicolumn{3}{c}{Age AME cover.\ \%} & \multicolumn{3}{c}{7 others: cover.\ \%} & \multicolumn{3}{c}{7 others: $|$bias$|$ (pp)} & \multicolumn{4}{c}{Test AUC} \\
\cmidrule(lr){4-6}\cmidrule(lr){7-9}\cmidrule(lr){10-12}\cmidrule(lr){13-16}
$p$ & $\beta_{\rm age}$ & $\rho$ & sp & B & s4b & sp & B & s4b & sp & B & s4b & sp & B & s4b & EN \\
\midrule
2000 & 0.5 & 0.5 & 98 & 70 & 99 & 96 & 23 & 97 & 1.3 & 3.7 & 1.3 & 0.915 & 0.907 & 0.900 & 0.822 \\
 & 0.5 & 0.8 & 98 & 23 & 100 & 97 & 21 & 97 & 1.3 & 3.8 & 1.3 & 0.916 & 0.905 & 0.900 & 0.822 \\
 & 1.0 & 0.5 & 99 & 80 & 98 & 96 & 19 & 96 & 1.3 & 3.6 & 1.3 & 0.928 & 0.921 & 0.914 & 0.845 \\
 & 1.0 & 0.8 & 93 & 51 & 100 & 95 & 21 & 95 & 1.3 & 3.6 & 1.3 & 0.926 & 0.916 & 0.913 & 0.842 \\
 & 1.5 & 0.5 & 93 & 77 & 98 & 96 & 15 & 96 & 1.3 & 3.4 & 1.3 & 0.939 & 0.933 & 0.928 & 0.867 \\
 & 1.5 & 0.8 & 95 & 59 & 99 & 96 & 16 & 97 & 1.2 & 3.4 & 1.2 & 0.939 & 0.930 & 0.928 & 0.870 \\
\addlinespace[2pt]
5000 & 0.5 & 0.5 & 97 & 63 & 99 & 95 & 20 & 96 & 1.4 & 3.9 & 1.4 & 0.908 & 0.898 & 0.890 & 0.820 \\
 & 0.5 & 0.8 & 97 & 21 & 99 & 96 & 18 & 97 & 1.3 & 4.0 & 1.4 & 0.906 & 0.896 & 0.888 & 0.821 \\
 & 1.0 & 0.5 & 97 & 77 & 96 & 95 & 18 & 97 & 1.3 & 3.7 & 1.3 & 0.917 & 0.906 & 0.902 & 0.838 \\
 & 1.0 & 0.8 & 92 & 49 & 100 & 95 & 17 & 95 & 1.3 & 3.7 & 1.4 & 0.918 & 0.905 & 0.903 & 0.839 \\
 & 1.5 & 0.5 & 89 & 83 & 99 & 96 & 13 & 97 & 1.2 & 3.5 & 1.3 & 0.932 & 0.916 & 0.918 & 0.865 \\
 & 1.5 & 0.8 & 93 & 55 & 100 & 95 & 14 & 96 & 1.3 & 3.4 & 1.3 & 0.929 & 0.916 & 0.918 & 0.866 \\
\bottomrule
\end{tabular}
}
\end{table}
\begin{table}[htbp]\centering
\caption{Study 1: covariate detection and latent coefficients, pooled over the 12 scenarios. BART selects covariates using the Bayesian FDR rule on their PIPs. AME detection is the proportion of 95\% intervals excluding zero. Mean/$\beta$ is the posterior mean divided by the true coefficient, and coverage refers to 95\% intervals on the latent scale.}
\label{tab:A1cov}
\resizebox{0.85\textwidth}{!}{
\begin{tabular}{lr c cc cc cc}
\toprule
& & & \multicolumn{2}{c}{AME CrI $\not\ni0$ (\%)} & \multicolumn{2}{c}{spBART $\beta$} & \multicolumn{2}{c}{\texttt{stan4bart} $\beta$} \\
\cmidrule(lr){4-5}\cmidrule(lr){6-7}\cmidrule(lr){8-9}
Covariate & $\beta$ & BART selects \% & sp & s4b & Mean/$\beta$ & Cover.\ \% & Mean/$\beta$ & Cover.\ \% \\
\midrule
age & 0.5--1.5 & 73 & 100 & 100 & 0.86 & 40 & 1.09 & 93 \\
sex & 0.30 & 15 & 81 & 76 & 0.86 & 93 & 1.09 & 96 \\
race & -0.30 & 8 & 71 & 67 & 0.84 & 94 & 1.07 & 96 \\
bmi & 0.25 & 29 & 100 & 99 & 0.86 & 89 & 1.09 & 95 \\
smoking & 0.20 & 2 & 36 & 33 & 0.87 & 94 & 1.11 & 96 \\
education & -0.15 & 2 & 78 & 74 & 0.85 & 95 & 1.08 & 96 \\
activity & -0.10 & 0 & 44 & 40 & 0.85 & 95 & 1.09 & 96 \\
famhx & 0.40 & 4 & 60 & 56 & 0.85 & 94 & 1.10 & 96 \\
\bottomrule
\end{tabular}
}
\end{table}

\subsection{Study 2: detecting a small covariate effect}\label{App::sim2}

We use DGM~2 with $n=1{,}500$, varying $\beta_{\mathrm{BMI}}$.
Table~\ref{tab:A2} reports the detection rates in
Figure~\ref{fig:sim2}, spBART's posterior coefficient means, and
interval coverage.

BART's covariate-PIP rule selects BMI in no replicates under
$\beta_{\mathrm{BMI}}=0$. To compare power at a nominal 5\%
false-positive rate, we use the 95th percentile of its null BMI PIP
as the threshold: 0.068 at $p=500$ and 0.009 at $p=2{,}000$.
For $\beta_{\mathrm{BMI}}=0.1$, 0.2, and 0.3, BART's detection rates
are 25\%, 74\%, and 100\% at $p=500$, and 9\%, 32\%, and 86\%
at $p=2{,}000$. The corresponding spBART rates are 42--48\%,
98--99\%, and 100\%. This oracle calibration relies on the simulated
null distribution and is unavailable in practice.

Averaged over the coefficient grid, spBART has gene sensitivity
0.999--1.000, empirical FDR 0.013--0.025, and test AUC 0.919--0.920.
The corresponding BART values are 0.998--1.000, 0.008--0.012,
and 0.917--0.918.

\begin{table}[htbp]\centering
\caption{Study 2: BMI detection rates (\%), spBART's posterior mean of $\beta_{\mathrm{BMI}}$, and coverage of its 95\% credible interval (\%). spBART detects an effect when the interval excludes zero. BART selects BMI using the Bayesian FDR rule on covariate PIPs. Each row summarizes 100 replicates.}
\label{tab:A2}
\resizebox{0.9\textwidth}{!}{
\begin{tabular}{r cccc cccc}
\toprule
& \multicolumn{4}{c}{$p=500$ (detection and coverage in \%)} & \multicolumn{4}{c}{$p=2000$} \\
\cmidrule(lr){2-5}\cmidrule(lr){6-9}
$\beta_{\rm BMI}$ & sp: CrI $\not\ni0$ & B: selects & sp mean & sp cover. & sp: CrI $\not\ni0$ & B: selects & sp mean & sp cover. \\
\midrule
0.00 & 4 & 0 & -0.002 & 96 & 5 & 0 & 0.005 & 95 \\
0.10 & 48 & 0 & 0.083 & 93 & 42 & 0 & 0.081 & 94 \\
0.20 & 98 & 30 & 0.167 & 90 & 99 & 16 & 0.170 & 92 \\
0.30 & 100 & 91 & 0.266 & 86 & 100 & 67 & 0.261 & 86 \\
0.45 & 100 & 100 & 0.395 & 78 & 100 & 100 & 0.393 & 79 \\
0.60 & 100 & 100 & 0.522 & 67 & 100 & 100 & 0.524 & 60 \\
0.80 & 100 & 100 & 0.698 & 51 & 100 & 100 & 0.699 & 50 \\
\bottomrule
\end{tabular}
}
\end{table}

\subsection{Study 3: semi-synthetic data built on the 5hmC matrix}\label{App::sim3}

In Study 3, we use the 115 screened genes from the MM application or 2,000 features drawn
once from the 14,147 filtered features. The application screen was
based on observed MM status and is kept fixed for the simulated
outcomes. Results for the larger set are conditional on its single
draw. In the clustered layout, the ten signals comprise a randomly
chosen gene and its nine most correlated neighbors.

The tree form of $f$ uses DGM~2 with signal-gene cutpoints at empirical
quantiles
\[
(0.45,0.55,0.40,0.60,0.50,0.50,0.35,0.65,0.42,0.58).
\]
The smooth form applies DGM~3 to the observed variance-stabilized
values. We use observed age (standardized), sex, race, and BMI
($\geq25$ vs.\ $<25$), with DGM~2 coefficients. Study membership
is excluded. All covariate effects are nonzero, so we assess power
and coverage. Study~2 reports covariate false-positive rates.

Table~\ref{tab:A5cov} compares detection by latent coefficient
intervals, AME intervals, and BART's covariate-PIP rule.
For spBART and \texttt{stan4bart}, the coefficient and AME intervals
give identical detection decisions. With screened genes, BART's
PIP rule selects sex, race, and BMI more often than spBART's intervals
detect them, while BART's AME detection and coverage are lower.
PIP selection therefore needs to be distinguished from effect
estimation. Tables~\ref{tab:A5sel} and \ref{tab:A5beta} report gene
selection for all five methods and latent coefficient summaries,
respectively.

\begin{table}[htbp]\centering
\caption{Study 3: covariate detection and AME interval coverage, both reported as percentages. Detection uses exclusion of zero from 95\% latent coefficient or AME intervals, or BART's covariate-PIP rule. Results are pooled over signal layout and form of $f$, with 800 datasets per gene set.}
\label{tab:A5cov}
\resizebox{0.9\textwidth}{!}{
\begin{tabular}{ll r cc c ccc ccc}
\toprule
& & & \multicolumn{2}{c}{$\beta$: 95\% CrI $\not\ni0$} & BART & \multicolumn{3}{c}{AME: 95\% CrI $\not\ni0$} & \multicolumn{3}{c}{AME coverage} \\
& & & & & selects & & & & & & \\
\cmidrule(lr){4-5}\cmidrule(lr){6-6}\cmidrule(lr){7-9}\cmidrule(lr){10-12}
Genes & Covariate & $\beta$ & sp & s4b & B & sp & B & s4b & sp & B & s4b \\
\midrule
115 screened & Age & 0.50 & 98 & 99 & 38 & 98 & 32 & 99 & 94 & 25 & 94 \\
 & Sex & 0.60 & 81 & 84 & 90 & 81 & 74 & 84 & 93 & 92 & 94 \\
 & Race & -0.40 & 18 & 20 & 51 & 18 & 6 & 20 & 95 & 76 & 96 \\
 & BMI & 0.45 & 51 & 57 & 79 & 51 & 43 & 57 & 95 & 89 & 95 \\
\addlinespace[2pt]
2000 random & Age & 0.50 & 94 & 94 & 20 & 94 & 19 & 94 & 94 & 17 & 96 \\
 & Sex & 0.60 & 75 & 75 & 45 & 75 & 43 & 75 & 94 & 54 & 95 \\
 & Race & -0.40 & 18 & 17 & 5 & 18 & 1 & 17 & 93 & 11 & 95 \\
 & BMI & 0.45 & 52 & 49 & 28 & 52 & 23 & 49 & 95 & 36 & 97 \\
\bottomrule
\end{tabular}
}
\end{table}
\begin{table}[htbp]\centering
\caption{Study 3: gene sensitivity, empirical FDR and AUC on the real validation set (knockoffs do not produce predictions). Each row summarizes 200 replicates.}
\label{tab:A5sel}
\resizebox{\textwidth}{!}{
\begin{tabular}{lll ccccc ccccc ccccc}
\toprule
& & & \multicolumn{5}{c}{Sensitivity} & \multicolumn{5}{c}{FDR} & \multicolumn{5}{c}{Test AUC (real validation set)} \\
\cmidrule(lr){4-8}\cmidrule(lr){9-13}\cmidrule(lr){14-18}
Genes & Signals & $f$ & sp & B & s4b & EN & KO & sp & B & s4b & EN & KO & sp & B & s4b & EN & KO \\
\midrule
115 screened & random & trees & 0.36 & 0.34 & 0.97 & 0.77 & 0.37 & 0.20 & 0.04 & 0.59 & 0.61 & 0.12 & 0.867 & 0.870 & 0.911 & 0.850 & -- \\
2000 random & random & trees & 0.20 & 0.17 & 0.21 & 0.51 & 0.18 & 0.40 & 0.38 & 0.01 & 0.83 & 0.16 & 0.788 & 0.768 & 0.849 & 0.784 & -- \\
115 screened & cluster & trees & 0.32 & 0.31 & 0.92 & 0.70 & 0.37 & 0.17 & 0.04 & 0.55 & 0.57 & 0.11 & 0.886 & 0.885 & 0.920 & 0.875 & -- \\
2000 random & cluster & trees & 0.22 & 0.22 & 0.11 & 0.77 & 0.19 & 0.32 & 0.29 & 0.00 & 0.81 & 0.11 & 0.914 & 0.906 & 0.934 & 0.918 & -- \\
115 screened & random & smooth & 0.21 & 0.20 & 0.56 & 0.31 & 0.23 & 0.27 & 0.06 & 0.61 & 0.65 & 0.22 & 0.796 & 0.795 & 0.820 & 0.771 & -- \\
2000 random & random & smooth & 0.12 & 0.11 & 0.09 & 0.22 & 0.15 & 0.47 & 0.37 & 0.01 & 0.80 & 0.20 & 0.759 & 0.742 & 0.785 & 0.739 & -- \\
115 screened & cluster & smooth & 0.18 & 0.17 & 0.57 & 0.30 & 0.21 & 0.29 & 0.04 & 0.57 & 0.55 & 0.24 & 0.804 & 0.804 & 0.826 & 0.762 & -- \\
2000 random & cluster & smooth & 0.10 & 0.09 & 0.06 & 0.25 & 0.11 & 0.49 & 0.39 & 0.01 & 0.71 & 0.32 & 0.802 & 0.784 & 0.823 & 0.762 & -- \\
\bottomrule
\end{tabular}
}
\end{table}
\begin{table}[htbp]\centering
\caption{Study 3: latent coefficients, posterior mean divided by the true value / coverage (\%) of the 95\% interval, for spBART and \texttt{stan4bart}.}
\label{tab:A5beta}
\resizebox{0.9\textwidth}{!}{
\begin{tabular}{lll cccc cccc}
\toprule
& & & \multicolumn{4}{c}{spBART: mean/$\beta$ / cover.\ \%} & \multicolumn{4}{c}{\texttt{stan4bart}: mean/$\beta$ / cover.\ \%} \\
\cmidrule(lr){4-7}\cmidrule(lr){8-11}
Genes & Signals & $f$ & Age & Sex & Race & BMI & Age & Sex & Race & BMI \\
\midrule
115 screened & random & trees & 0.62/28 & 0.66/78 & 0.68/95 & 0.64/84 & 1.05/93 & 1.12/94 & 1.09/96 & 1.10/94 \\
2000 random & random & trees & 0.61/25 & 0.61/60 & 0.59/86 & 0.65/80 & 0.90/94 & 0.86/98 & 0.89/94 & 0.93/94 \\
115 screened & cluster & trees & 0.58/26 & 0.59/70 & 0.58/88 & 0.62/85 & 0.95/94 & 0.97/96 & 0.96/98 & 1.02/97 \\
2000 random & cluster & trees & 0.51/24 & 0.52/64 & 0.42/85 & 0.51/79 & 0.84/94 & 0.84/97 & 0.77/94 & 0.87/97 \\
115 screened & random & smooth & 0.67/38 & 0.71/75 & 0.62/90 & 0.70/81 & 0.88/88 & 0.91/91 & 0.91/94 & 0.92/94 \\
2000 random & random & smooth & 0.65/32 & 0.68/68 & 0.60/86 & 0.70/86 & 0.83/84 & 0.85/93 & 0.79/94 & 0.88/97 \\
115 screened & cluster & smooth & 0.70/44 & 0.79/84 & 0.69/90 & 0.77/88 & 0.90/88 & 0.98/96 & 0.97/93 & 0.96/94 \\
2000 random & cluster & smooth & 0.72/54 & 0.80/85 & 0.71/89 & 0.74/88 & 0.96/94 & 1.04/94 & 0.97/92 & 0.96/96 \\
\bottomrule
\end{tabular}
}
\end{table}

\subsection{Study 4: covariates acting only through tree interactions}\label{App::sim4}

This study compares spBART and BART using the selection procedure of Section~\ref{Sec::24} under DGM~1, where covariates act only through gene--covariate interactions, and DGM~2, where they act linearly. DGM~1 misspecifies spBART's linear term, whereas DGM~2 matches its specification. We use $n\in\{1{,}500,2{,}500\}$, $p\in\{500,2{,}000,3{,}000\}$ and 500 replicates per setting. For both methods, we use the same five cross-validation folds, apply the Bayesian FDR rule within each fold, retain the union of selected genes, and refit on this union before test-set evaluation. We exclude BART's covariate PIPs from the gene FDR rule. We also fit unpenalized logistic and probit regressions with all genes and four covariates as linear terms.

\begin{figure}[htbp]\centering
\includegraphics[width=0.9\textwidth]{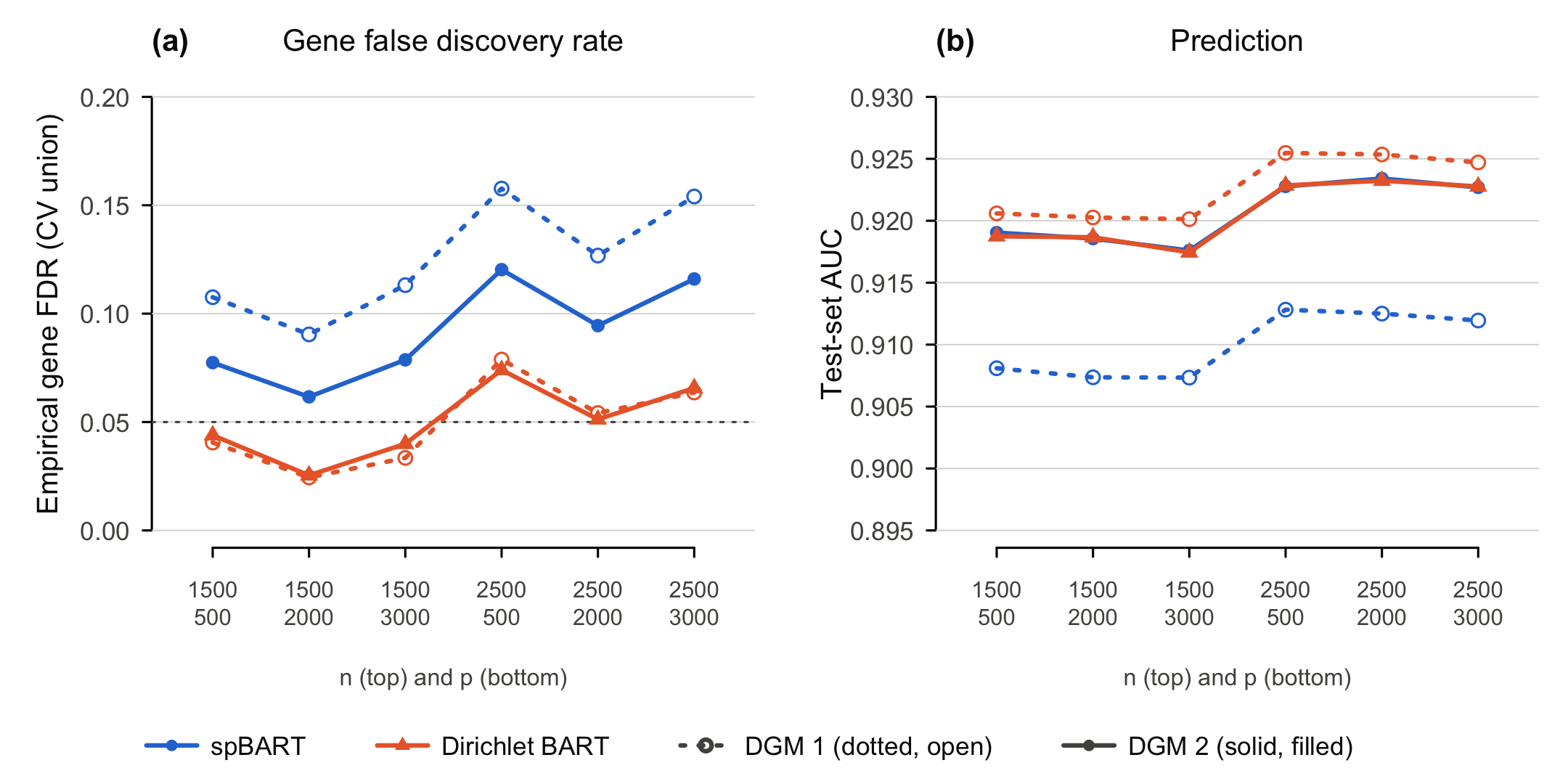}
\caption{Study 4: empirical FDR of the cross-validation union (dotted line: 0.05) and test-set AUC by $n$ (top) and $p$ (bottom).}
\label{fig:A8}
\end{figure}

We recover all ten signal genes with both methods in every setting (Table~\ref{tab:A8sel}). The fold union also retains null genes, giving empirical FDR of 0.06--0.16 for spBART and 0.02--0.08 for BART. With spBART, we select an average of 0.4--1.3 additional genes, all null. Under DGM~2, we obtain comparable test AUCs of 0.917--0.923, with differences of at most 0.001. Under DGM~1, we lose 0.012--0.013 in AUC with spBART relative to BART (Table~\ref{tab:A8pred}, Figure~\ref{fig:A8}). The interaction-only design therefore favors BART in prediction, while we retain the same gene recovery with spBART.

Under DGM~2, we select each covariate with BART in most datasets, with race selected least often (54--70\% at $n=1{,}500$). With spBART, we estimate coefficients at 0.87--0.94 of their true values. We obtain below-nominal coverage for age (75--81\%) and BMI (78--83\%) and near-nominal coverage for race (93--96\%) (Table~\ref{tab:A8cov}). We thus obtain interpretable coefficient summaries alongside comparable prediction, although attenuation limits inference on effect size. The unpenalized logistic and probit fits exhibit separation in every dataset at $p\ge2{,}000$, with separation rates of 13--37\% at $n=1{,}500$, $p=500$, so we cannot use these fits as reliable coefficient benchmarks.

\begin{table}[htbp]\centering
\caption{Study 4: gene selection with the cross-validation union procedure: sensitivity, empirical FDR and mean number of selected genes. Each row summarizes 500 replicates.}
\label{tab:A8sel}
\resizebox{\textwidth}{!}{
\begin{tabular}{rr ccc ccc ccc ccc}
\toprule
& & \multicolumn{6}{c}{DGM 1 (trees only; spBART misspecified)} & \multicolumn{6}{c}{DGM 2 (trees + linear $Z\beta$)} \\
\cmidrule(lr){3-8}\cmidrule(lr){9-14}
& & \multicolumn{3}{c}{spBART} & \multicolumn{3}{c}{BART} & \multicolumn{3}{c}{spBART} & \multicolumn{3}{c}{BART} \\
\cmidrule(lr){3-5}\cmidrule(lr){6-8}\cmidrule(lr){9-11}\cmidrule(lr){12-14}
$n$ & $p$ & Sens. & FDR & \# sel. & Sens. & FDR & \# sel. & Sens. & FDR & \# sel. & Sens. & FDR & \# sel. \\
\midrule
1500 & 500 & 1.00 & 0.11 & 11.3 & 1.00 & 0.04 & 10.5 & 1.00 & 0.08 & 10.9 & 1.00 & 0.04 & 10.5 \\
1500 & 2000 & 1.00 & 0.09 & 11.1 & 1.00 & 0.02 & 10.3 & 1.00 & 0.06 & 10.7 & 1.00 & 0.03 & 10.3 \\
1500 & 3000 & 1.00 & 0.11 & 11.4 & 1.00 & 0.03 & 10.4 & 1.00 & 0.08 & 10.9 & 1.00 & 0.04 & 10.5 \\
2500 & 500 & 1.00 & 0.16 & 12.0 & 1.00 & 0.08 & 10.9 & 1.00 & 0.12 & 11.5 & 1.00 & 0.07 & 10.9 \\
2500 & 2000 & 1.00 & 0.13 & 11.6 & 1.00 & 0.05 & 10.6 & 1.00 & 0.09 & 11.1 & 1.00 & 0.05 & 10.6 \\
2500 & 3000 & 1.00 & 0.15 & 12.0 & 1.00 & 0.06 & 10.7 & 1.00 & 0.12 & 11.4 & 1.00 & 0.07 & 10.8 \\
\bottomrule
\end{tabular}
}
\end{table}
\begin{table}[htbp]\centering
\caption{Study 4: test-set AUC and Brier score of the final model fitted on the union of selected genes.}
\label{tab:A8pred}
\resizebox{0.8\textwidth}{!}{
\begin{tabular}{rr cccc cccc}
\toprule
& & \multicolumn{4}{c}{DGM 1} & \multicolumn{4}{c}{DGM 2} \\
\cmidrule(lr){3-6}\cmidrule(lr){7-10}
& & \multicolumn{2}{c}{spBART} & \multicolumn{2}{c}{BART} & \multicolumn{2}{c}{spBART} & \multicolumn{2}{c}{BART} \\
\cmidrule(lr){3-4}\cmidrule(lr){5-6}\cmidrule(lr){7-8}\cmidrule(lr){9-10}
$n$ & $p$ & AUC & Brier & AUC & Brier & AUC & Brier & AUC & Brier \\
\midrule
1500 & 500 & 0.908 & 0.123 & 0.921 & 0.115 & 0.919 & 0.116 & 0.919 & 0.117 \\
1500 & 2000 & 0.907 & 0.124 & 0.920 & 0.116 & 0.919 & 0.116 & 0.919 & 0.117 \\
1500 & 3000 & 0.907 & 0.124 & 0.920 & 0.116 & 0.918 & 0.117 & 0.917 & 0.118 \\
2500 & 500 & 0.913 & 0.120 & 0.925 & 0.111 & 0.923 & 0.113 & 0.923 & 0.113 \\
2500 & 2000 & 0.913 & 0.120 & 0.925 & 0.111 & 0.923 & 0.112 & 0.923 & 0.113 \\
2500 & 3000 & 0.912 & 0.121 & 0.925 & 0.112 & 0.923 & 0.113 & 0.923 & 0.113 \\
\bottomrule
\end{tabular}
}
\end{table}
\begin{table}[htbp]\centering
\caption{Study 4, DGM~2: BART covariate selection (\%, Bayesian FDR rule on the covariate PIPs) and spBART's latent coefficients (posterior mean divided by the true value; coverage \% of 95\% intervals).}
\label{tab:A8cov}
\resizebox{\textwidth}{!}{
\begin{tabular}{rr ccc ccc ccc ccc}
\toprule
& & \multicolumn{3}{c}{Age} & \multicolumn{3}{c}{Sex} & \multicolumn{3}{c}{Race} & \multicolumn{3}{c}{BMI} \\
\cmidrule(lr){3-5}\cmidrule(lr){6-8}\cmidrule(lr){9-11}\cmidrule(lr){12-14}
$n$ & $p$ & B sel.\ \% & sp mean/$\beta$ & sp cover.\ \% & B sel.\ \% & sp mean/$\beta$ & sp cover.\ \% & B sel.\ \% & sp mean/$\beta$ & sp cover.\ \% & B sel.\ \% & sp mean/$\beta$ & sp cover.\ \% \\
\midrule
1500 & 500 & 100 & 0.88 & 75 & 100 & 0.87 & 87 & 70 & 0.88 & 93 & 100 & 0.88 & 78 \\
1500 & 2000 & 100 & 0.88 & 78 & 99 & 0.88 & 89 & 54 & 0.88 & 94 & 100 & 0.88 & 80 \\
1500 & 3000 & 100 & 0.88 & 75 & 98 & 0.89 & 88 & 54 & 0.90 & 95 & 100 & 0.88 & 80 \\
2500 & 500 & 100 & 0.92 & 80 & 100 & 0.91 & 89 & 95 & 0.91 & 94 & 100 & 0.91 & 81 \\
2500 & 2000 & 100 & 0.91 & 79 & 100 & 0.92 & 90 & 88 & 0.93 & 93 & 100 & 0.91 & 83 \\
2500 & 3000 & 100 & 0.92 & 81 & 100 & 0.91 & 89 & 90 & 0.94 & 96 & 100 & 0.91 & 83 \\
\bottomrule
\end{tabular}
}
\end{table}

\subsection{Study 5: very high dimension without screening}\label{App::sim5}

\begin{figure}[htbp]
\centering
\includegraphics[width=\textwidth]{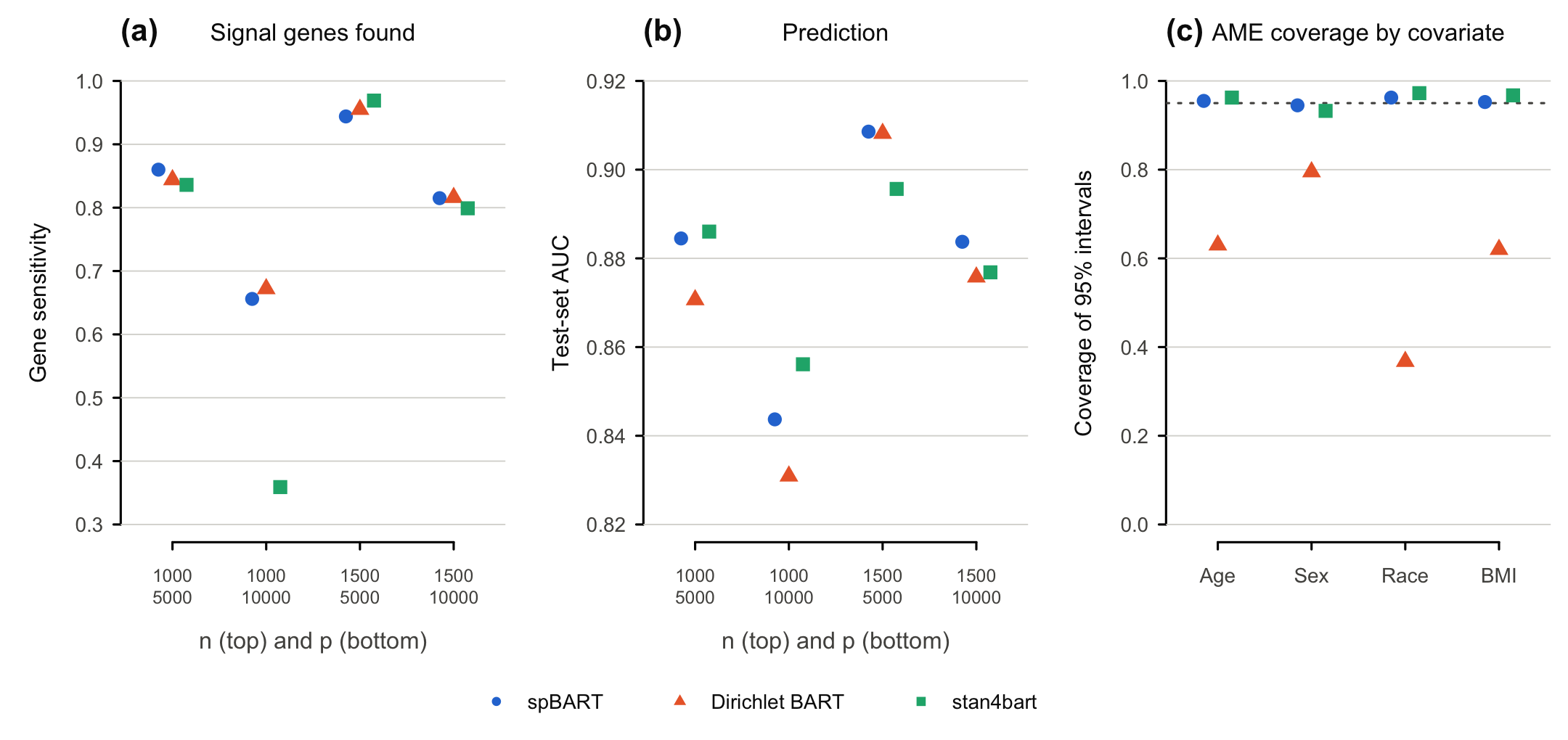}
\caption{Gene selection, prediction, and covariate inference in
Study~5, based on 100 replicates per setting. Panels show
(a) the proportion of ten signal genes selected, (b) test-set AUC,
and (c) coverage of 95\% AME intervals for the four covariates,
averaged over settings. In (a) and (b), axis labels give $n$ above
$p$. The dotted line in (c) marks 0.95.}
\label{fig:sim3}
\end{figure}

The application screens 14{,}147 genes down to 115 before fitting the model (Section~\ref{Sec::41}). Study~5 asks how the methods perform without screening. We use DGM~2, with ten signal genes acting through shallow trees and linear effects of age, sex, race and BMI, and consider $p\in\{5{,}000,10{,}000\}$ and $n\in\{1{,}000,1{,}500\}$, with 100 replicates in each of the four settings.

spBART and Dirichlet BART have similar gene sensitivity, 0.66--0.94 and 0.67--0.95, respectively, with empirical gene FDRs of 0.02--0.07 for both methods (Figure~\ref{fig:sim3}a and Table~\ref{tab:A3sel}). At $n=1{,}000$ and $p=10{,}000$, \texttt{stan4bart} selects 36\% of the signal genes. The elastic net selects every signal gene, but with an empirical FDR of 0.67--0.70 and an AUC of 0.80--0.83. The test AUC of spBART exceeds that of Dirichlet BART by 0.008--0.014 in three settings and is essentially unchanged in the fourth, with paired Monte Carlo standard errors of 0.002--0.004. \texttt{stan4bart} has the highest AUC at $n=1{,}000$.

The methods differ more in covariate inference (Figure~\ref{fig:sim3}c and Table~\ref{tab:A3cov}). The coverage of the 95\% AME intervals is 92--98\% for spBART and 29--91\% for Dirichlet BART, whose lowest coverage is for race, a binary covariate with 30\% prevalence that Dirichlet BART selects in only 10--32\% of datasets. The latent coefficients of spBART are attenuated, at 0.70--0.86 of their true values, whereas those of \texttt{stan4bart} are less biased (0.85--1.08). With independent genes and at least 1{,}000 participants, spBART can therefore be fitted without screening, with gene selection and prediction comparable to those of Dirichlet BART, although its sensitivity falls to 0.66 in the hardest setting. Gene detection is much harder with the correlated real genes and $n=500$ (Study~3), so the value of screening in the application remains an open question.

\begin{table}[htbp]\centering
\caption{Study 5: gene sensitivity, empirical FDR and test-set AUC. Each row summarizes 100 replicates.}
\label{tab:A3sel}
\resizebox{0.85\textwidth}{!}{
\begin{tabular}{rr cccc cccc cccc}
\toprule
& & \multicolumn{4}{c}{Sensitivity} & \multicolumn{4}{c}{FDR} & \multicolumn{4}{c}{Test AUC} \\
\cmidrule(lr){3-6}\cmidrule(lr){7-10}\cmidrule(lr){11-14}
$n$ & $p$ & sp & B & s4b & EN & sp & B & s4b & EN & sp & B & s4b & EN \\
\midrule
1000 & 5000 & 0.86 & 0.84 & 0.84 & 1.00 & 0.04 & 0.02 & 0.03 & 0.67 & 0.884 & 0.871 & 0.886 & 0.807 \\
1000 & 10000 & 0.66 & 0.67 & 0.36 & 1.00 & 0.07 & 0.04 & 0.00 & 0.67 & 0.844 & 0.831 & 0.856 & 0.804 \\
1500 & 5000 & 0.94 & 0.95 & 0.97 & 1.00 & 0.02 & 0.02 & 0.09 & 0.67 & 0.909 & 0.908 & 0.896 & 0.828 \\
1500 & 10000 & 0.82 & 0.82 & 0.80 & 1.00 & 0.05 & 0.05 & 0.01 & 0.70 & 0.884 & 0.876 & 0.877 & 0.823 \\
\bottomrule
\end{tabular}
}
\end{table}
\begin{table}[htbp]\centering
\caption{Study 5: covariate inference. BART selection rates and coverage of 95\% AME intervals are reported as percentages. For spBART and \texttt{stan4bart}, coefficient summaries give the posterior mean divided by the true coefficient and coverage (\%) of 95\% intervals on the latent scale.}
\label{tab:A3cov}
\resizebox{0.8\textwidth}{!}{
\begin{tabular}{rrl c ccc cc cc}
\toprule
& & & & \multicolumn{3}{c}{AME cover.\ \%} & \multicolumn{2}{c}{spBART $\beta$} & \multicolumn{2}{c}{\texttt{stan4bart} $\beta$} \\
\cmidrule(lr){5-7}\cmidrule(lr){8-9}\cmidrule(lr){10-11}
$n$ & $p$ &  & B selects \% & sp & B & s4b & Mean/$\beta$ & Cover. & Mean/$\beta$ & Cover. \\
\midrule
1000 & 5000 & Age & 71 & 92 & 55 & 95 & 0.77 & 53 & 1.05 & 98 \\
 &  & Sex & 63 & 93 & 75 & 93 & 0.79 & 76 & 1.08 & 93 \\
 &  & Race & 13 & 97 & 29 & 99 & 0.76 & 92 & 1.02 & 100 \\
 &  & BMI & 69 & 95 & 58 & 97 & 0.78 & 54 & 1.06 & 97 \\
\addlinespace[2pt]
1000 & 10000 & Age & 54 & 97 & 48 & 98 & 0.71 & 26 & 0.87 & 97 \\
 &  & Sex & 54 & 95 & 66 & 93 & 0.70 & 56 & 0.88 & 95 \\
 &  & Race & 10 & 96 & 36 & 100 & 0.74 & 83 & 0.92 & 98 \\
 &  & BMI & 52 & 96 & 49 & 97 & 0.72 & 33 & 0.89 & 97 \\
\addlinespace[2pt]
1500 & 5000 & Age & 99 & 98 & 76 & 96 & 0.84 & 63 & 1.01 & 97 \\
 &  & Sex & 89 & 96 & 91 & 95 & 0.85 & 89 & 1.01 & 96 \\
 &  & Race & 32 & 96 & 51 & 97 & 0.83 & 91 & 1.00 & 96 \\
 &  & BMI & 92 & 94 & 78 & 97 & 0.86 & 69 & 1.04 & 94 \\
\addlinespace[2pt]
1500 & 10000 & Age & 87 & 95 & 73 & 96 & 0.79 & 47 & 0.89 & 93 \\
 &  & Sex & 86 & 94 & 86 & 92 & 0.79 & 71 & 0.89 & 89 \\
 &  & Race & 18 & 96 & 31 & 93 & 0.75 & 88 & 0.85 & 94 \\
 &  & BMI & 76 & 96 & 63 & 96 & 0.79 & 45 & 0.89 & 93 \\
\bottomrule
\end{tabular}
}
\end{table}

\subsection{Study 6: sensitivity to the prior variance of beta and the contributions of the two components}\label{App::sim6}

Study~6 assesses sensitivity to the prior variance $\sigma_\beta^2\in\{0.1,1,10,100,1{,}000\}$ in three scenarios with 100 replicates each. Two scenarios use DGM~2 with $n=1{,}500$ and $p\in\{500,2{,}000\}$. The semi-synthetic scenario uses the real development-set predictors ($n=500$, the 115 screened genes and the four covariates), with age standardized and the other covariates binary (Table~\ref{Table21}). In this scenario we choose ten signal genes at random, generate outcomes with the DGM~2 gene trees, empirical-quantile cutpoints and covariate coefficients, and test on the validation-set predictors ($n=369$). To see how much each component contributes to prediction, we also compute test AUCs from the linear term alone, $\Phi(\mathbf{Z}^\top\boldsymbol{\beta})$, and from the tree ensemble alone, $\Phi(f)$, of the same fit. At each posterior draw, let $V_Z$ and $V_f$ be the empirical variances across subjects of $\mathbf{Z}_i^\top\boldsymbol{\beta}$ and $f(\widetilde{\mathbf{X}}_i)$. We report the posterior means of the normalized variance shares
\[
\frac{V_Z}{V_Z+V_f+1},
\qquad
\frac{V_f}{V_Z+V_f+1},
\]
where the remaining share $1/(V_Z+V_f+1)$ belongs to the unit-variance latent error.

\begin{figure}[htbp]
\centering
\includegraphics[width=\textwidth]{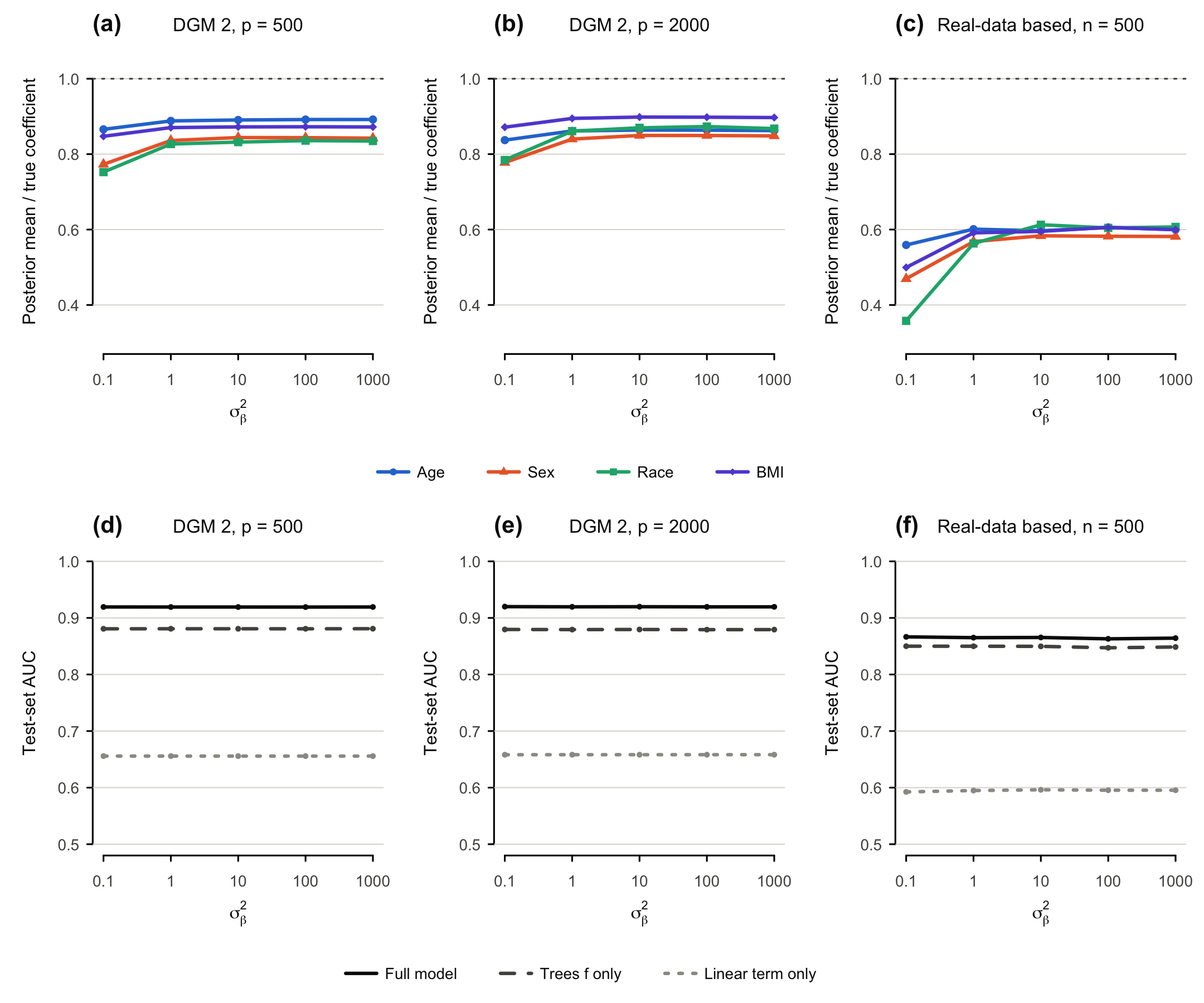}
\caption{Prior sensitivity and component contributions in Study~6,
with $\sigma_\beta^2$ shown on a log scale. The top row shows posterior
coefficient means divided by their true values. The bottom row shows
test-set AUCs using the full model and each component of the same fit.
Columns represent DGM~2 with $p=500$ and $p=2{,}000$ ($n=1{,}500$),
and the semi-synthetic scenario ($n=500$). Each point
summarizes 100 replicates.}
\label{fig:sim4}
\end{figure}

Gene selection, AUC and the coefficient estimates vary little for $\sigma_\beta^2\geq1$, a range that includes our default of 10, whereas $\sigma_\beta^2=0.1$ shrinks the coefficients further (Figure~\ref{fig:sim4} and Table~\ref{tab:A4}). In DGM~2, the AUCs are approximately 0.66 for the linear term, 0.88 for the trees and 0.92 for the full model. The corresponding values in the semi-synthetic scenario are 0.60, 0.85 and 0.87. Most of the discrimination therefore comes from the genes, and the covariates add to it. The variance share of the linear component is approximately 0.13 in DGM~2. In the semi-synthetic scenario, sensitivity is approximately 0.36 and the empirical FDR approximately 0.19 for every value of $\sigma_\beta^2$, consistent with the difficulty of gene selection in Study~3.

The coefficients remain attenuated under diffuse priors, with posterior means of approximately 0.83--0.90 of the true values in DGM~2 and 0.60 in the semi-synthetic scenario. In Studies~1, 3 and 5, the ratio of the posterior mean to the true coefficient is generally closer to one for \texttt{stan4bart} (0.77--1.12) than for spBART (0.42--0.87). The probability-scale AMEs are less affected in these studies, which supports reporting them alongside the latent-scale coefficients.

\begin{table}[htbp]\centering
\caption{Study 6: gene selection, test-set AUCs for the full model and its linear and tree components, normalized posterior variance shares, and latent coefficient summaries by $\sigma_\beta^2$. Coefficient summaries give the posterior mean divided by the true value and coverage of 95\% intervals, both averaged over the four covariates. Each row summarizes 100 replicates.}
\label{tab:A4}
\resizebox{0.95\textwidth}{!}{
\begin{tabular}{lr cc ccc cc cc}
\toprule
& & \multicolumn{2}{c}{Genes} & \multicolumn{3}{c}{Test AUC} & \multicolumn{2}{c}{Variance share} & \multicolumn{2}{c}{$\bm\beta$ (mean of 4)} \\
\cmidrule(lr){3-4}\cmidrule(lr){5-7}\cmidrule(lr){8-9}\cmidrule(lr){10-11}
Scenario & $\sigma^2_\beta$ & Sens. & FDR & full & $Z\beta$ only & $f$ only & $Z\beta$ & $f$ & $\hat\beta/\beta$ & Cover.\ \% \\
\midrule
DGM 2, $p{=}500$ & 0.1 & 1.00 & 0.02 & 0.919 & 0.656 & 0.881 & 0.12 & 0.60 & 0.81 & 74 \\
 & 1 & 1.00 & 0.02 & 0.919 & 0.656 & 0.881 & 0.12 & 0.60 & 0.86 & 85 \\
 & 10 & 1.00 & 0.03 & 0.919 & 0.656 & 0.881 & 0.13 & 0.60 & 0.86 & 86 \\
 & 100 & 1.00 & 0.02 & 0.919 & 0.656 & 0.881 & 0.13 & 0.60 & 0.86 & 86 \\
 & 1000 & 1.00 & 0.03 & 0.919 & 0.656 & 0.881 & 0.13 & 0.60 & 0.86 & 86 \\
\addlinespace[2pt]
DGM 2, $p{=}2000$ & 0.1 & 1.00 & 0.01 & 0.920 & 0.658 & 0.880 & 0.12 & 0.60 & 0.82 & 72 \\
 & 1 & 1.00 & 0.02 & 0.920 & 0.658 & 0.879 & 0.12 & 0.60 & 0.86 & 82 \\
 & 10 & 1.00 & 0.02 & 0.920 & 0.658 & 0.880 & 0.13 & 0.60 & 0.87 & 84 \\
 & 100 & 1.00 & 0.02 & 0.920 & 0.658 & 0.879 & 0.13 & 0.59 & 0.87 & 84 \\
 & 1000 & 1.00 & 0.02 & 0.920 & 0.658 & 0.879 & 0.13 & 0.59 & 0.87 & 83 \\
\addlinespace[2pt]
Semi-syn., $n{=}500$ & 0.1 & 0.36 & 0.20 & 0.867 & 0.592 & 0.850 & 0.06 & 0.48 & 0.47 & 45 \\
 & 1 & 0.37 & 0.18 & 0.865 & 0.595 & 0.850 & 0.08 & 0.48 & 0.58 & 62 \\
 & 10 & 0.36 & 0.19 & 0.865 & 0.596 & 0.850 & 0.08 & 0.48 & 0.60 & 62 \\
 & 100 & 0.36 & 0.19 & 0.863 & 0.595 & 0.847 & 0.08 & 0.48 & 0.60 & 64 \\
 & 1000 & 0.36 & 0.21 & 0.864 & 0.595 & 0.849 & 0.08 & 0.48 & 0.60 & 62 \\
\bottomrule
\end{tabular}
}
\end{table}

\subsection{Study 7: one long chain versus many short chains}\label{App::sim7}

In this study, we compare five chain strategies for the full training-set fits of
spBART and BART, outlined below:
\begin{itemize}
    \item S0: uses one chain of 7,000 iterations.
    \item S10-eq: uses ten chains of 700 iterations, with 200 burn-in iterations per chain, matching S0's total iteration budget.
    \item S10, S20, and S50: use 10, 20, and 50 chains of 3,000 iterations, respectively, with 1,000 burn-in iterations per chain and pooled posterior draws.
\end{itemize}
Each chain uses a different seed, with spBART initialized at
$\boldsymbol{\beta}^{(0)}\sim\mathcal{N}(\mathbf{0},4I)$.

The study scenarios include DGMs~1--2 with $n=1{,}500$ and
$p\in\{500,2{,}000\}$, DGM~2 with covariates strongly correlated
with signal genes ($c=0.9$, $p=500$, as in Study~8), and the semi-synthetic scenario
of Study~6. We analyze 100 data replicates for each scenario. We measure agreement with S0
by the Jaccard index of selected gene sets and compute rank-normalized
split-$\widehat{R}$ \citep{vehtari2021rank} for $\boldsymbol{\beta}$,
log-likelihood, split count, and $f$ at a test point. S0 diagnostics
use the two halves of its single chain.

\begin{figure}[htbp]\centering
\includegraphics[width=0.9\textwidth]{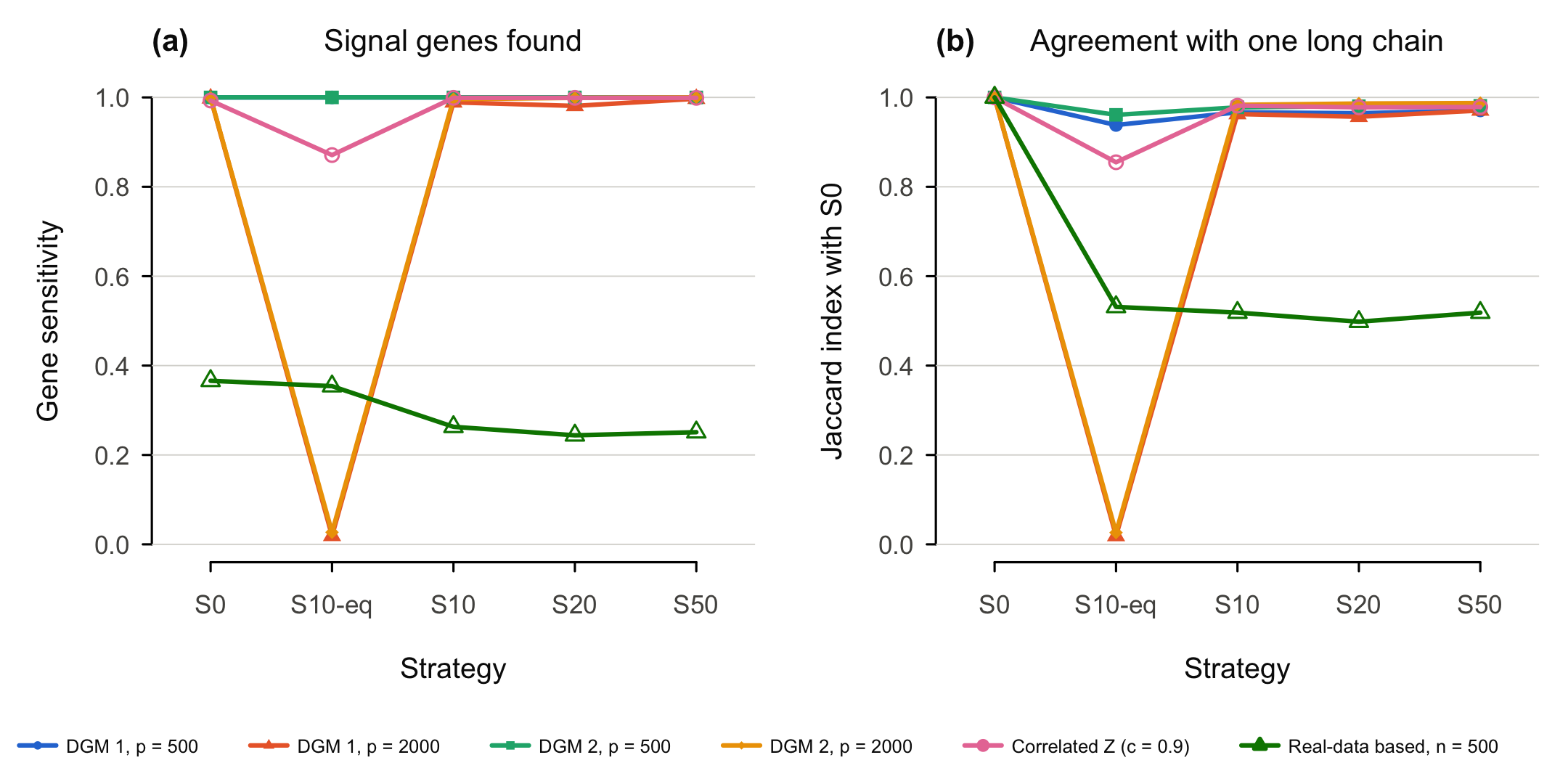}
\caption{Study 7: spBART sensitivity and agreement (Jaccard index) with the single long chain S0, by strategy and scenario.}
\label{fig:A6}
\end{figure}

In the fully simulated scenarios, S10-eq recovers only 2--3\% of
signals at $p=2{,}000$ for both methods
(Table~\ref{tab:A6} and Figure~\ref{fig:A6}). The longer S10 and S50
chains give high agreement with S0 (Jaccard 0.96--0.99), similar
test AUC, and slightly lower FDR. Additional chains therefore add
little to selection or prediction in these settings.

The diagnostics show incomplete mixing despite this agreement.
The largest coefficient $\widehat{R}$ is approximately 1.01 at
$p=500$, 1.05--1.07 at $p=2{,}000$, and 1.17 when covariates strongly
correlate with signal genes. Study~8 finds coefficient bias under strong correlation
at $p=2{,}000$, whereas this chain comparison uses $p=500$.
Coefficient $\widehat{R}$ also exceeds 1.01 in most independent-covariate
replicates at $p=2{,}000$, so that threshold does not distinguish
the correlated setting. Log-likelihood $\widehat{R}$ of 1.5--1.7
at $p=2{,}000$ indicates incomplete exploration of tree space
(Table~\ref{tab:A6rhat}). 

The application-sized scenario shows less agreement (Jaccard
approximately 0.5). For spBART, S10 and S50 reduce FDR from 0.19
to 0.01--0.02 and sensitivity from 0.37 to 0.25--0.26, with AUC
increasing from approximately 0.87 to 0.88. BART shows the same
pattern. This comparison supports selecting from pooled draws of
adequately long chains in application-sized analyses, while recognizing
the loss in sensitivity. Our application uses the single long-chain strategy, and Appendix~\ref{App::multichain} checks the eight selected loci across independently initialized chains fitted to the 115 screened genes.

\begin{table}[htbp]\centering
\caption{Study 7: gene selection, prediction and agreement with the single long chain (S0), by sampling strategy. The Jaccard index measures agreement between each strategy's selected gene set and S0's. Max $\widehat{R}_\beta$ is the largest split-$\widehat{R}$ among the four coefficients, with the percentage of replicates exceeding 1.01 also reported. Each row summarizes 100 replicates.}
\label{tab:A6}
\resizebox{0.85\textwidth}{!}{
\begin{tabular}{ll ccc c cc ccc}
\toprule
& & \multicolumn{6}{c}{spBART} & \multicolumn{3}{c}{BART} \\
\cmidrule(lr){3-8}\cmidrule(lr){9-11}
Scenario & Strategy & Sens. & FDR & AUC & Jaccard & max $\hat R_\beta$ & \% $>$1.01 & Sens. & FDR & Jaccard \\
\midrule
DGM 1, $p{=}500$ & S0 & 1.00 & 0.04 & 0.909 & -- & 1.01 & 27 & 1.00 & 0.01 & -- \\
 & S10-eq & 1.00 & 0.05 & 0.908 & 0.94 & 1.04 & 100 & 1.00 & 0.03 & 0.97 \\
 & S10 & 1.00 & 0.03 & 0.909 & 0.97 & 1.01 & 54 & 1.00 & 0.01 & 0.99 \\
 & S50 & 1.00 & 0.02 & 0.910 & 0.97 & 1.01 & 47 & 1.00 & 0.01 & 0.99 \\
\addlinespace[1pt]
DGM 1, $p{=}2000$ & S0 & 1.00 & 0.02 & 0.908 & -- & 1.01 & 27 & 1.00 & 0.01 & -- \\
 & S10-eq & 0.02 & 0.00 & 0.880 & 0.02 & 1.30 & 100 & 0.02 & 0.00 & 0.02 \\
 & S10 & 0.99 & 0.00 & 0.907 & 0.96 & 1.07 & 96 & 0.98 & 0.00 & 0.97 \\
 & S50 & 1.00 & 0.00 & 0.908 & 0.97 & 1.07 & 100 & 1.00 & 0.00 & 0.99 \\
\addlinespace[1pt]
DGM 2, $p{=}500$ & S0 & 1.00 & 0.02 & 0.921 & -- & 1.01 & 26 & 1.00 & 0.01 & -- \\
 & S10-eq & 1.00 & 0.04 & 0.919 & 0.96 & 1.04 & 100 & 1.00 & 0.02 & 0.97 \\
 & S10 & 1.00 & 0.01 & 0.921 & 0.98 & 1.01 & 21 & 1.00 & 0.01 & 0.99 \\
 & S50 & 1.00 & 0.01 & 0.921 & 0.98 & 1.01 & 7 & 1.00 & 0.01 & 0.99 \\
\addlinespace[1pt]
DGM 2, $p{=}2000$ & S0 & 1.00 & 0.01 & 0.919 & -- & 1.01 & 14 & 1.00 & 0.01 & -- \\
 & S10-eq & 0.03 & 0.00 & 0.893 & 0.03 & 1.30 & 100 & 0.02 & 0.00 & 0.02 \\
 & S10 & 1.00 & 0.00 & 0.918 & 0.98 & 1.05 & 89 & 0.99 & 0.00 & 0.98 \\
 & S50 & 1.00 & 0.00 & 0.918 & 0.99 & 1.06 & 100 & 1.00 & 0.00 & 0.99 \\
\addlinespace[1pt]
Corr.\ $Z$ $c{=}0.9$ & S0 & 0.99 & 0.02 & 0.932 & -- & 1.05 & 73 & 1.00 & 0.01 & -- \\
 & S10-eq & 0.87 & 0.00 & 0.926 & 0.85 & 1.61 & 100 & 0.98 & 0.00 & 0.97 \\
 & S10 & 1.00 & 0.01 & 0.932 & 0.98 & 1.17 & 100 & 1.00 & 0.00 & 0.99 \\
 & S50 & 1.00 & 0.00 & 0.932 & 0.98 & 1.17 & 100 & 1.00 & 0.00 & 0.99 \\
\addlinespace[1pt]
Semi-syn., $n{=}500$ & S0 & 0.37 & 0.19 & 0.867 & -- & 1.02 & 63 & 0.35 & 0.04 & -- \\
 & S10-eq & 0.35 & 0.01 & 0.891 & 0.53 & 1.09 & 100 & 0.29 & 0.01 & 0.57 \\
 & S10 & 0.26 & 0.02 & 0.881 & 0.52 & 1.04 & 100 & 0.27 & 0.01 & 0.61 \\
 & S50 & 0.25 & 0.01 & 0.882 & 0.52 & 1.04 & 100 & 0.24 & 0.00 & 0.60 \\
\bottomrule
\end{tabular}
}
\end{table}
\begin{table}[htbp]\centering
\caption{Study 7: spBART between-chain $\widehat{R}$ for the log-likelihood, the number of splits and $f$ at a fixed test point, for 10 and 50 chains of 3{,}000 iterations.}
\label{tab:A6rhat}
\resizebox{0.65\textwidth}{!}{
\begin{tabular}{l ccc ccc}
\toprule
& \multicolumn{3}{c}{S10: $\hat R$} & \multicolumn{3}{c}{S50: $\hat R$} \\
\cmidrule(lr){2-4}\cmidrule(lr){5-7}
Scenario & log-lik. & \# splits & $f$(test 1) & log-lik. & \# splits & $f$(test 1) \\
\midrule
DGM 1, $p{=}500$ & 1.11 & 1.01 & 1.04 & 1.10 & 1.01 & 1.04 \\
DGM 1, $p{=}2000$ & 1.66 & 1.01 & 1.15 & 1.68 & 1.01 & 1.15 \\
DGM 2, $p{=}500$ & 1.07 & 1.01 & 1.03 & 1.08 & 1.01 & 1.03 \\
DGM 2, $p{=}2000$ & 1.52 & 1.01 & 1.14 & 1.52 & 1.01 & 1.14 \\
Corr.\ $Z$ $c{=}0.9$ & 1.17 & 1.01 & 1.05 & 1.19 & 1.01 & 1.06 \\
Semi-syn., $n{=}500$ & 1.37 & 1.07 & 1.13 & 1.40 & 1.08 & 1.13 \\
\bottomrule
\end{tabular}
}
\end{table}

\subsection{Study 8: smooth, correlated and partially misspecified data-generating models}\label{App::sim8}

In this study, we compare spBART, BART, \texttt{stan4bart} and the elastic net under DGM~2 and four departures from it, using $n=1{,}500$, $p\in\{500,2{,}000\}$ and 100 replicates for each setting. In addition, we do the following:
\begin{itemize}
\item We use DGM~3 to assess a smooth gene effect.
\item We generate blocks of 50 correlated genes as $X_{ij}=\sqrt{\rho}\,W_{i,b(j)}+\sqrt{1-\rho}\,\varepsilon_{ij}$, with $\rho\in\{0.5,0.8\}$. We place each signal gene in a separate block or five signals in each of two blocks. We also evaluate block-level selection, counting a selected block as a discovery if it contains a signal gene.
\item We generate gene--covariate dependence through
\[
\begin{aligned}
\text{age} &= cX_1+\sqrt{1-c^2}\,e_1, &
\text{BMI} &= cX_2+\sqrt{1-c^2}\,e_4,\\
\text{sex} &= \mathbb{I}\{cX_3+\sqrt{1-c^2}\,e_2>0\}, &
\text{race} &= \mathbb{I}\{cX_4+\sqrt{1-c^2}\,e_3>\Phi^{-1}(0.7)\}.
\end{aligned}
\]
We vary $c\in\{0,0.3,0.6,0.9\}$ and use either the DGM~2 trees or a near-linear gene effect, $0.565\sum_{j=1}^{10}X_j$, with the same variance. We examine coefficient estimation when the genes strongly predict $\mathbf{Z}$ \citep{robinson1988root}.
\item We introduce nonlinear age and BMI effects,
\[
0.4\tanh(1.5\,\text{age})+0.3(\text{BMI}^2-1),
\]
and two DGM~1 gene--covariate interaction trees involving genes 1 and 3. We retain linear sex and race effects of $0.6$ and $-0.4$. To assess partial misspecification, we fit spBART with age and BMI in the linear term or in the trees.
\end{itemize}

Under DGM~2, we obtain similar results for spBART and BART, with sensitivity 1.00, FDR 0.01--0.02 and test AUC 0.917--0.920 (Table~\ref{tab:A7main}). We observe higher FDR for \texttt{stan4bart} (0.13--0.26) and the elastic net (0.58--0.64), whose AUC is about 0.08 lower. With smooth $f$, we recover roughly half the signals using the tree-based methods and 30\% using the elastic net. Several smooth components have weak marginal structure. We obtain the highest AUC with spBART (0.857--0.862).

With strongly correlated genes ($\rho=0.8$), we obtain similar gene-level sensitivity for spBART and BART, falling to 0.73--0.91 in the harder layouts. For both methods, we retain block-level sensitivity of at least 0.99 and block-level FDR of at most 0.02 (Table~\ref{tab:A7corrX}). We therefore recover the signal blocks even when inclusion probabilities spread across correlated neighbors.

\begin{figure}[htbp]\centering
\includegraphics[width=0.95\textwidth]{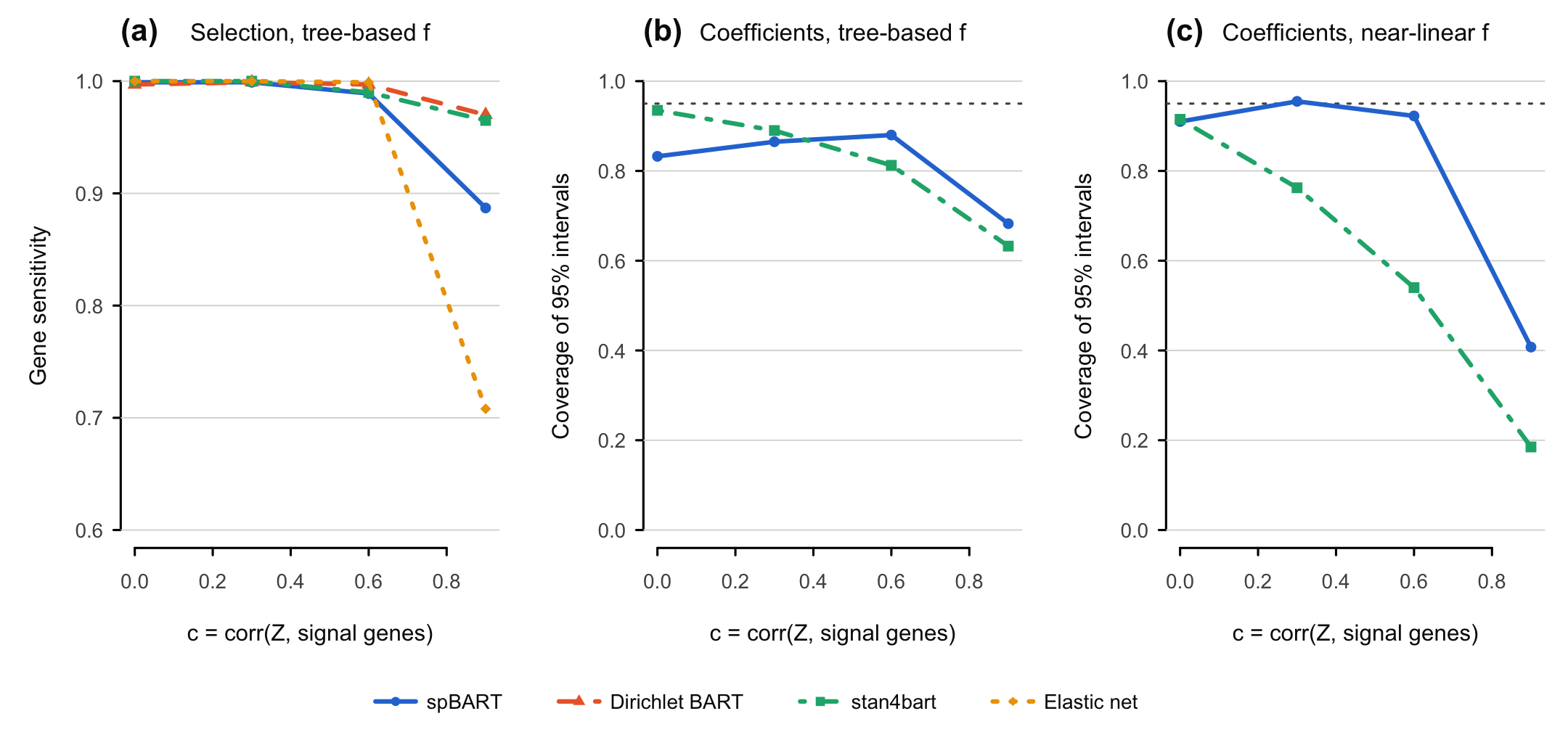}
\caption{Study 8: effect of the correlation $c$ between the covariates and the signal genes ($p=2{,}000$). Gene sensitivity with tree-based $f$, and mean coverage of the 95\% intervals for the four coefficients with tree-based and near-linear $f$.}
\label{fig:A7}
\end{figure}

With gene--covariate dependence, we observe little change in spBART's selection and age, sex and BMI coefficients through $c=0.6$, although the race coefficient falls to 0.70--0.74 of its true value. At $c=0.9$, we obtain sensitivity of 0.78--0.89 for spBART and 0.96--0.97 for BART. With near-linear $f$, we overestimate the age and BMI coefficients by factors of 1.7--1.9 with spBART and 2.1--2.2 with \texttt{stan4bart}, with poor coverage for spBART and near-zero coverage for \texttt{stan4bart} (Table~\ref{tab:A7corrZ}, Figure~\ref{fig:A7}). These biases are consistent with regularization-induced confounding \citep{hahn2018regularization}. In Study~7, we also observe larger between-chain $\widehat{R}$ under strong gene--covariate dependence.

Under partial misspecification, we lose 0.007--0.009 in AUC relative to BART when we place nonlinear age and BMI effects in the linear term. Moving these covariates into the trees restores comparable AUC and improves interval coverage for sex and race, although \texttt{stan4bart} estimates these linear effects most accurately (Table~\ref{tab:A7partial}). Across most scenarios, we obtain lower FDR with spBART than with \texttt{stan4bart} and the elastic net. With strongly correlated genes at $p=2{,}000$, however, \texttt{stan4bart} has lower FDR. We thus retain comparable prediction and selection in several departures from DGM~2, while strong gene--covariate dependence limits coefficient inference and nonlinear covariate effects require a more flexible adjustment.

\begin{table}[htbp]\centering
\caption{Study 8: DGM~2 baseline, smooth $f$ (DGM~3) and partial misspecification: gene sensitivity, empirical FDR and test-set AUC. Each row summarizes 100 replicates.}
\label{tab:A7main}
\resizebox{\textwidth}{!}{
\begin{tabular}{lr cccc cccc cccc}
\toprule
& & \multicolumn{4}{c}{Sensitivity} & \multicolumn{4}{c}{FDR} & \multicolumn{4}{c}{Test AUC} \\
\cmidrule(lr){3-6}\cmidrule(lr){7-10}\cmidrule(lr){11-14}
Scenario & $p$ & sp & B & s4b & EN & sp & B & s4b & EN & sp & B & s4b & EN \\
\midrule
DGM 2 baseline & 500 & 1.00 & 1.00 & 1.00 & 1.00 & 0.02 & 0.02 & 0.26 & 0.58 & 0.920 & 0.919 & 0.913 & 0.836 \\
 & 2000 & 1.00 & 1.00 & 1.00 & 1.00 & 0.02 & 0.01 & 0.13 & 0.64 & 0.919 & 0.917 & 0.906 & 0.831 \\
Smooth $f$ & 500 & 0.59 & 0.58 & 0.60 & 0.30 & 0.03 & 0.01 & 0.27 & 0.32 & 0.862 & 0.859 & 0.841 & 0.797 \\
 & 2000 & 0.53 & 0.51 & 0.55 & 0.30 & 0.06 & 0.03 & 0.12 & 0.35 & 0.857 & 0.852 & 0.835 & 0.793 \\
Partial misspec. & 500 & 1.00 & 1.00 & 1.00 & 1.00 & 0.02 & 0.01 & 0.25 & 0.57 & 0.915 & 0.924 & 0.909 & 0.830 \\
 & 2000 & 1.00 & 1.00 & 1.00 & 1.00 & 0.02 & 0.01 & 0.13 & 0.64 & 0.915 & 0.922 & 0.901 & 0.826 \\
\bottomrule
\end{tabular}
}
\end{table}
\begin{table}[htbp]\centering
\caption{Study 8: gene-level and block-level sensitivity and FDR under block-correlated genes. Each signal gene is placed in a separate block ("separate") or five signal genes in each of two blocks ("same block").}
\label{tab:A7corrX}
\resizebox{\textwidth}{!}{
\begin{tabular}{lr ccc ccc ccc ccc}
\toprule
& & \multicolumn{3}{c}{Sensitivity} & \multicolumn{3}{c}{FDR} & \multicolumn{3}{c}{Block sens.} & \multicolumn{3}{c}{Block FDR} \\
\cmidrule(lr){3-5}\cmidrule(lr){6-8}\cmidrule(lr){9-11}\cmidrule(lr){12-14}
Scenario & $p$ & sp & B & s4b & sp & B & s4b & sp & B & s4b & sp & B & s4b \\
\midrule
$\rho{=}0.5$, separate & 500 & 1.00 & 1.00 & 1.00 & 0.02 & 0.01 & 0.22 & 1.00 & 1.00 & 1.00 & 0.00 & 0.00 & 0.00 \\
 & 2000 & 0.99 & 0.99 & 1.00 & 0.04 & 0.01 & 0.10 & 1.00 & 1.00 & 1.00 & 0.01 & 0.00 & 0.06 \\
$\rho{=}0.8$, separate & 500 & 1.00 & 1.00 & 1.00 & 0.03 & 0.01 & 0.18 & 1.00 & 1.00 & 1.00 & 0.00 & 0.00 & 0.00 \\
 & 2000 & 0.84 & 0.82 & 0.97 & 0.17 & 0.18 & 0.04 & 0.99 & 0.99 & 0.98 & 0.00 & 0.00 & 0.01 \\
$\rho{=}0.8$, same block & 500 & 0.91 & 0.91 & 0.99 & 0.02 & 0.01 & 0.12 & 1.00 & 1.00 & 1.00 & 0.01 & 0.00 & 0.19 \\
 & 2000 & 0.73 & 0.74 & 0.79 & 0.08 & 0.08 & 0.04 & 1.00 & 1.00 & 1.00 & 0.02 & 0.00 & 0.01 \\
\bottomrule
\end{tabular}
}
\end{table}
\begin{table}[htbp]\centering
\caption{Study 8: gene selection and coefficient estimation under gene--covariate dependence, with $p=2{,}000$. Dependence is varied through $c$ and uses tree-based or near-linear gene effects. Gene sensitivity is reported for spBART and BART, and posterior means divided by the true coefficients and coverage (\%) of 95\% intervals for spBART and \texttt{stan4bart}.}
\label{tab:A7corrZ}
\resizebox{\textwidth}{!}{
\begin{tabular}{l cc cccc cccc}
\toprule
& \multicolumn{2}{c}{Sensitivity} & \multicolumn{4}{c}{spBART: mean/$\beta$ / cover.\ \%} & \multicolumn{4}{c}{\texttt{stan4bart}: mean/$\beta$ / cover.\ \%} \\
\cmidrule(lr){2-3}\cmidrule(lr){4-7}\cmidrule(lr){8-11}
$f$, $c$ & sp & B & Age & Sex & Race & BMI & Age & Sex & Race & BMI \\
\midrule
trees, $c{=}0$ & 1.00 & 1.00 & 0.87/74 & 0.88/90 & 0.85/94 & 0.85/75 & 1.13/90 & 1.15/92 & 1.10/98 & 1.10/94 \\
trees, $c{=}0.3$ & 1.00 & 1.00 & 0.89/81 & 0.86/87 & 0.85/91 & 0.90/87 & 1.19/82 & 1.15/94 & 1.07/96 & 1.20/84 \\
trees, $c{=}0.6$ & 0.99 & 1.00 & 0.91/85 & 0.93/92 & 0.74/88 & 0.92/87 & 1.25/75 & 1.30/79 & 0.84/97 & 1.26/74 \\
trees, $c{=}0.9$ & 0.89 & 0.97 & 1.19/61 & 1.16/77 & 0.34/68 & 1.15/67 & 1.46/48 & 1.53/66 & 0.44/88 & 1.49/51 \\
\addlinespace[2pt]
linear, $c{=}0$ & 1.00 & 1.00 & 0.91/86 & 0.89/95 & 0.88/96 & 0.91/87 & 1.15/86 & 1.13/96 & 1.13/98 & 1.14/86 \\
linear, $c{=}0.3$ & 1.00 & 1.00 & 0.95/93 & 0.98/100 & 0.82/93 & 0.97/96 & 1.27/65 & 1.32/75 & 0.94/96 & 1.28/69 \\
linear, $c{=}0.6$ & 0.99 & 1.00 & 1.02/96 & 1.00/97 & 0.70/81 & 1.03/95 & 1.41/34 & 1.44/59 & 0.62/84 & 1.45/39 \\
linear, $c{=}0.9$ & 0.78 & 0.96 & 1.72/11 & 1.31/71 & 0.43/77 & 1.85/4 & 2.12/0 & 1.84/30 & -0.12/44 & 2.24/0 \\
\bottomrule
\end{tabular}
}
\end{table}
\begin{table}[htbp]\centering
\caption{Study 8: partial misspecification. Latent coefficients of the linearly acting covariates (posterior mean divided by the true value, coverage \%), gene sensitivity and test-set AUC.}
\label{tab:A7partial}
\resizebox{0.75\textwidth}{!}{
\begin{tabular}{rl cc cc cc}
\toprule
& & \multicolumn{2}{c}{Sex ($\beta=0.6$)} & \multicolumn{2}{c}{Race ($\beta=-0.4$)} & & \\
\cmidrule(lr){3-4}\cmidrule(lr){5-6}
$p$ & Model & Mean/$\beta$ & Cover.\ \% & Mean/$\beta$ & Cover.\ \% & Sens. & AUC \\
\midrule
500 & spBART (all 4 in $Z$) & 0.81 & 76 & 0.75 & 89 & 1.00 & 0.915 \\
 & spBART (age, BMI in trees) & 0.82 & 83 & 0.77 & 92 & 1.00 & 0.925 \\
 & \texttt{stan4bart} & 1.08 & 95 & 1.00 & 99 & 1.00 & 0.909 \\
 & BART & -- & -- & -- & -- & 1.00 & 0.924 \\
2000 & spBART (all 4 in $Z$) & 0.78 & 72 & 0.82 & 89 & 1.00 & 0.915 \\
 & spBART (age, BMI in trees) & 0.79 & 80 & 0.81 & 90 & 1.00 & 0.922 \\
 & \texttt{stan4bart} & 1.01 & 97 & 1.07 & 97 & 1.00 & 0.901 \\
 & BART & -- & -- & -- & -- & 1.00 & 0.922 \\
\bottomrule
\end{tabular}
}
\end{table}


\section{Multi-chain convergence and selection stability in the MM application}\label{App::multichain}

We repeat fold-wise selection, the union and refitting with four MCMC chains per fit. To do this, we retain the primary analysis's 115 screened genes, study indicator, folds, priors and fold-specific FDR level of $\alpha=0.05$. Each cross-validation fit uses 400 participants with age and sex in the linear component. The refit uses all 500 development participants and all four covariates. Each chain has a separate seed and random latent-variable initialization, with 10{,}000 burn-in iterations followed by 10{,}000 iterations thinned every tenth. We retain 1{,}000 draws per chain and pool 4{,}000 draws per fit.

We assess rank-normalized split-$\widehat{R}$ and bulk effective sample size (ESS) \citep{vehtari2021rank} using thresholds of $\widehat{R}\le1.01$ and ESS above 400. The coefficients, split counts and monitored latent means meet both criteria in every fit (Table~\ref{tab:MC_diag1}). All monitored quantities also meet these criteria in the final development-set refit, supporting the posterior summaries reported for that model. Some cross-validation log-likelihood diagnostics indicate slower mixing, so we also examine gene selection across individual chains.

Gene selection varies across chains, particularly for genes near the FDR threshold (Tables~\ref{tab:MC_gene}). \textit{IL1RAP} appears in every chain--fold fit, while \textit{ST5} and \textit{NF2} appear in every chain's union. The pooled union contains six genes and the study indicator. All seven variables have PIP 1.00 in each refit chain, but these PIPs describe their use after selection. They do not establish that we would recover the same set from the 115 candidates.

\begin{table}[htbp]
\centering
\caption{Convergence diagnostics across four chains. We report split-$\widehat{R}$ and bulk ESS for the log-likelihood and split count, and the largest $\widehat{R}$ and smallest ESS among the coefficients and among three fixed participants' latent means. Cross-validation fits include age and sex coefficients, while the refit includes all four covariates.}
\label{tab:MC_diag1}
\resizebox{\textwidth}{!}{%
\begin{tabular}{l cc cc cc cc}
\toprule
& \multicolumn{2}{c}{Log-likelihood} & \multicolumn{2}{c}{Number of splits} & \multicolumn{2}{c}{Coefficients} & \multicolumn{2}{c}{Latent means} \\
\cmidrule(lr){2-3}\cmidrule(lr){4-5}\cmidrule(lr){6-7}\cmidrule(lr){8-9}
Fit & $\widehat{R}$ & ESS & $\widehat{R}$ & ESS & max $\widehat{R}$ & min ESS & max $\widehat{R}$ & min ESS \\
\midrule
CV fold 1 & 1.004 & 1507 & 1.000 & 3527 & 1.003 & 703 & 1.003 & 2212 \\
CV fold 2 & 1.022 & 233 & 1.002 & 3743 & 1.004 & 688 & 1.002 & 2434 \\
CV fold 3 & 1.022 & 274 & 1.004 & 3750 & 1.008 & 674 & 1.001 & 2200 \\
CV fold 4 & 1.004 & 2316 & 1.000 & 3602 & 1.005 & 632 & 1.003 & 2240 \\
CV fold 5 & 1.016 & 946 & 1.002 & 3588 & 1.007 & 700 & 1.003 & 2116 \\
Refit (development set) & 1.000 & 3126 & 1.000 & 3758 & 1.005 & 1111 & 1.001 & 2612 \\
\bottomrule
\end{tabular}}
\end{table}

\begin{table}[htbp]
\centering
\caption{Selection across folds and chains for variables entering any union. Primary refers to the original single-chain analysis. Chain--fold counts summarize the 20 individual fits, and chain unions count the chains selecting a variable in at least one fold. Pooled results use draws combined across four chains. Mean fold PIP averages the pooled PIPs across folds, while the PIP range covers individual chains and folds.}
\label{tab:MC_gene}
\resizebox{\textwidth}{!}{%
\begin{tabular}{l c c c c c c c}
\toprule
Variable & Primary (of 5) & Chain--fold fits (of 20) & Chain unions (of 4) & Pooled folds (of 5) & In pooled union & Mean fold PIP & PIP range \\
\midrule
Study indicator & 5 & 20 & 4 & 5 & yes & 1.00 & 1.00--1.00 \\
\textit{IL1RAP} & 5 & 20 & 4 & 5 & yes & 0.97 & 0.94--0.99 \\
\textit{ST5} & 4 & 13 & 4 & 5 & yes & 0.92 & 0.90--0.94 \\
\textit{KL} & 1 & 6 & 3 & 1 & yes & 0.90 & 0.86--0.94 \\
\textit{HERC6} & 2 & 2 & 1 & 0 & no & 0.89 & 0.85--0.93 \\
\textit{UBR4} & 1 & 2 & 2 & 0 & no & 0.86 & 0.82--0.93 \\
\textit{MYO1E} & 2 & 1 & 1 & 0 & no & 0.87 & 0.82--0.92 \\
\textit{ELK3} & 1 & 1 & 1 & 0 & no & 0.89 & 0.83--0.94 \\
\textit{CAPN2} & 1 & 1 & 1 & 0 & no & 0.80 & 0.74--0.92 \\
\addlinespace
\textit{NF2} & 0 & 11 & 4 & 2 & yes & 0.92 & 0.89--0.95 \\
\textit{SRGAP2} & 0 & 3 & 3 & 1 & yes & 0.85 & 0.80--0.94 \\
\textit{PRDM10} & 0 & 3 & 3 & 1 & yes & 0.85 & 0.80--0.93 \\
\textit{CARD11} & 0 & 2 & 2 & 0 & no & 0.79 & 0.73--0.92 \\
\textit{KIF15} & 0 & 1 & 1 & 0 & no & 0.89 & 0.83--0.93 \\
\bottomrule
\end{tabular}}
\end{table}

\section{Gene selection in the BC-only analysis}\label{App::BConly}

This section provides details on the analysis using BC participants only (Section~\ref{Sec::41}). We randomly split 576 BC participants into 500 development and 76 validation subjects, independently of the pooled split. We repeat screening, five-fold selection at $\alpha=0.05$, the fold union and refitting, with age, sex, race and BMI in the linear component and no study indicator. The folds select 7, 10, 7, 9 and 10 genes, giving a union of 19. Table~\ref{tab:BConly_gene} compares these selections with the primary and multi-chain pooled analyses.

\begin{table}[htbp]\centering
\caption{Gene selection in the BC-only and pooled analyses. Each entry reports the number of folds selecting each gene in the BC-only and primary pooled analyses, and the number of pooled chain unions containing it. ``Not screened'' indicates failure to pass BC-only screening. The genes are ordered by BC-only fold count and remaining primary pooled genes are listed in the lower block.}
\label{tab:BConly_gene}
\begin{tabular}{lccc}
\toprule
Gene & BC-only folds (of 5) & Pooled primary (of 5) & Pooled chain unions (of 4) \\
\midrule
\textit{IL1RAP} & 5 & 5 & 4 \\
\textit{FKBP9} & 5 & 0 & 0 \\
\textit{NF2} & 4 & 0 & 4 \\
\textit{PRKDC} & 4 & 0 & 0 \\
\textit{PKP2} & 3 & 0 & 0 \\
\textit{KCNMA1} & 3 & 0 & 0 \\
\textit{MSI2} & 3 & 0 & 0 \\
\textit{ST5} & 2 & 4 & 4 \\
\textit{PRDM10} & 2 & 0 & 3 \\
\textit{STARD13} & 2 & 0 & 0 \\
\textit{SRGAP3} & 2 & 0 & 0 \\
\textit{CARD11} & 1 & 0 & 2 \\
\textit{CIRH1A} & 1 & 0 & 0 \\
\textit{SEMA5B} & 1 & 0 & 0 \\
\textit{TLN2} & 1 & 0 & 0 \\
\textit{PPFIA1} & 1 & 0 & 0 \\
\textit{FSTL4} & 1 & 0 & 0 \\
\textit{LRFN2} & 1 & 0 & 0 \\
\textit{ARNT2} & 1 & 0 & 0 \\
\addlinespace
\textit{MYO1E} & 0 & 2 & 1 \\
\textit{UBR4} & 0 & 1 & 2 \\
\textit{ELK3} & 0 & 1 & 1 \\
\textit{CAPN2} & 0 & 1 & 1 \\
\textit{HERC6} & not screened & 2 & 1 \\
\textit{KL} & not screened & 1 & 3 \\
\bottomrule
\end{tabular}
\end{table}

\begin{figure}[htbp]\centering
\includegraphics[width=0.85\textwidth]{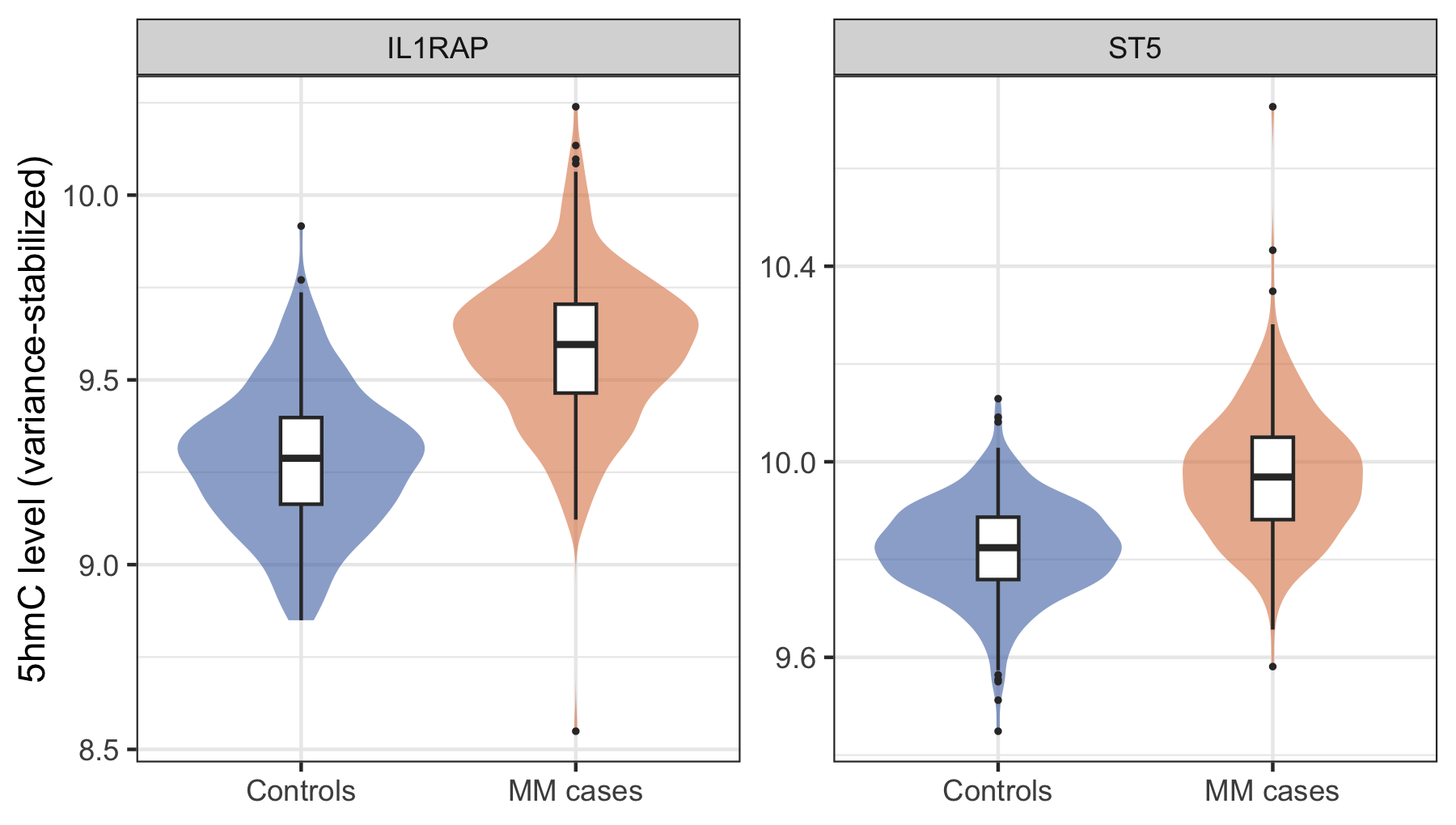}
\caption{Distributions of the 5hmC levels (variance-stabilized) of \textit{IL1RAP} and \textit{ST5} among the 282 MM cases and 294 controls of the BC case-control. Violins show the density, and boxes show the median and interquartile range. }
\label{fig:IL1RAP_ST5_BC}
\end{figure}

\end{document}